\documentclass[preprint]{aastex}

\def\cm2{\,{\rm cm}^{-2} }

\def\kmsm{\,{\rm km\, s^{-1}\, Mpc^{-1}} }
\def\keV{\,{\rm keV} }

\def\ergcms{\,{\rm erg\, cm^{-2}\, s^{-1}} }

\def\ergs{\,{\rm erg\, s^{-1}}}
\def\ergsh2{\,{\rm erg\, s^{-1}}, h=\frac{1}{2} }

\def\msun{\,{\rm M}_\odot }

\begin{document}

\title{Mass Models and Sunyaev-Zeldovich Effect Predictions 
\\ for a Flux Limited Sample of 22 Nearby X-Ray Clusters}

\author{Brian S. Mason \altaffilmark{1} and Steven T. Myers \altaffilmark{2}}
\affil{University of Pennsylvania \\ 209 S. 33rd St. Philadelphia, PA 19104-6396}

\altaffiltext{1}{\footnotesize Current Address:  California Institute of Technology, 105-24,  Pasadena, CA 91125}

\altaffiltext{2}{\footnotesize Current Address: NRAO, Socorro, NM 87801}

\begin{abstract}

We define a 90\% complete, volume-limited sample of 31 $z < 0.1$ x-ray
clusters and present a systematic analysis of public ROSAT PSPC data
on 22 of these objects. Our efforts are undertaken in support of the
Penn/OVRO SZE survey, and to this end we present predictions for the
inverse Compton optical depth towards all 22 of these clusters.  We
have performed detailed Monte Carlo simulations to understand the
effects of the cluster profile uncertainties on the SZE predictions
given the OVRO 5.5-meter telescope beam and switching patterns. We
also present a similar analysis for the near-future ACBAR experiment.
For most of the clusters in the sample we find less than a 5\%
uncertainty in the SZE predictions due to an imperfect knowledge of
the profile.  A comparison of different cooling flow modeling
strategies shows that our results are robust with respect to this.
The profile uncertainties are then one of the least significant
components of our error budget for SZE-based distance measurements.
The density models which result from this analysis also yield baryonic
masses and, under the assumption of hydrostatic equilibrium, total
masses and baryon mass fractions.  Our Monte Carlo profile analysis
indicates that the baryon masses within $1 \, h_{100}^{-1}\,$ Mpc for
these clusters are accurate to better than $\sim 5\%$ and unaffected
by realistic PSPC systematics.  In the sample as a whole, we find a
mean gas mass fraction of $(7.02 \pm 0.28) \, h_{100}^{-3/2} \, \times
10^{-2}$ internal to $R_{500} \sim 1 \, h_{100}^{-1} \,$ Mpc.  This is in
agreement with previous x-ray cluster analyses, which indicate an
overabundance of baryons relative to the prediction of Big Bang
Nucleosynthesis for an $\Omega_M = 1$ universe.  Our analysis of the
x-ray spectra confirms a previous claim of an excess absorbing column
density towards A478, but we do not find evidence for anomalous column
densities in the other 21 clusters.  We also find some indications of
an excess of soft counts in the ROSAT PSPC data.

A measurement of $H_o$ using these models and OVRO SZE determinations
will be presented in a second paper.

\end{abstract}


\section{Introduction}
\label{sec:introduction}

The discovery \citep{Giacconi_et_al_1972} that many galaxy clusters
are also strong sources of x-rays opened a new window on cosmology
which has proved fruitful for nearly three decades.  Subsequent
investigations
\citep[e.g.,][]{Mitchell_et_al_1976,Bahcall_and_Sarazin_1977}
determined that this emission originates in a hot ($kT_e \sim 7 \keV$)
thermal plasma with electron number densities of $\sim {\rm \, a \,
few} \times 10^{-3} {\rm \, cm^{-3}}$.  Since the sound-crossing time
for pressure waves in this plasma is less than the Hubble time, the
plasma can be assumed to accurately trace the cluster gravitational
potential.  Cluster virial masses obtained in this manner, in
conjunction with intracluster medium (ICM) models derived from
Einstein and ROSAT observations, have shown an overdensity of baryons
relative to the expectation for an $\Omega_M=1$, Big Bang
Nucleosynthesis universe
\citep{White_et_al_1995,White_and_Fabian_1995,Mohr_et_al_1999}, thus
providing a powerful challenge to the cosmological orthodoxy.

The inverse Compton scattering of Cosmic Microwave Background (CMB)
photons--- known as the Sunyaev-Zeldovich Effect (SZE)--- provides
another direct probe of the ICM.  As first indicated by
\citet{Cavaliere_et_al_1979}, the combination of x-ray and SZE
measurements on a given cluster yield a {\it direct} measurement of
the distance to the cluster.  If the thermal SZE decrement predicted
by the x-ray data under the assumption $h=1$ is $\Delta T_{pred}$,
and the observed decrement is $\Delta T_{obs}$, then $h$ is given by
\begin{equation}
h = \left(\frac{\Delta T_{pred}}{\Delta T_{obs}}\right)^2.
\end{equation}
Such a calculation requires knowledge of the structure of the cluster
along the line of sight; since the x-ray data do not directly provide
such information, the clusters are typically assumed to be spherically
symmetric.  It then becomes important to select clusters from an
orientation-unbiased sample.  Since the error in $H_o$ is twice the
error in $\Delta T_{pred}$, it is also important to accurately
understand the statistical and systematic uncertainties inherent in
the x-ray models.

The objective of the Penn/OVRO SZE survey is to determine $H_o$ from
observations of the SZE in an x-ray flux-limited sample of nearby
clusters.  The OVRO 5.5-meter telescope is an ideal instrument for
this purpose.  At 32 GHz, this telescope has a primary beam of $7'.35$
(FWHM) and a dual-horn switching angle of $22'.16$.  At the mean
redshift of our sample, these correspond to $ 425 \, h^{-1} \, {\rm
kpc}$ and $1.25\, h^{-1} \, {\rm Mpc}$, respectively.  Since the gas
in clusters is distributed on a scale $r_{core} \sim 200 \, h^{-1} \,
{\rm kpc}$, and most of the gas is contained within the inner $\sim 1
h^{-1}\, {\rm Mpc}$, the 5.5-meter main beam samples an
astrophysically relevant scale while the switching only removes
$5-10\%$ of the signal.  The first results from this survey were
reported in \citet{Myers_et_al_1997}:   with SZE measurements of four
clusters (Coma, Abell 478, Abell 2142, and Abell 2256) and x-ray
models from the literature, Myers et al. find $H_o = 54 \pm 14 \kmsm$.
The accuracy of these results is limited primarily by the x-ray
models.  

With this in mind, we have undertaken to expand the sample of Myers et
al.  and, using public x-ray data, construct high-quality x-ray models
and rigorously evaluate their reliability, taking into account the
specific observing strategy employed by our instrument. The spatial
resolution and large field of view of the ROSAT PSPC make it ideal for
our purposes.  While the PSPC does not have the energy or spatial
resolution of Chandra or XMM, it has a larger field of view than any
instrument on either of these observatories and so imaging analyses of
extended objects based on ROSAT data will continue to be relevant for
the foreseeable future.

The density models which result from our analysis will also be useful
for near-future experiments capable of measuring the SZ distortion
resulting from the bulk motion of the cluster gas.  To this end we
present predictions for the beam-averaged optical depth for the
near-future ACBAR experiment, a sensitive bolometric receiver which
will begin taking observations in Antarctica early in 2001.  This
instrument has four frequency channels between 150 GHz and 345 GHz,
with matched $4'$ beams, and will be sensitive to both the thermal and
kinematic SZ effects.

In the following section, we first present a brief summary of ICM
models (\S~\ref{sec:icmmodels}), along with the formalism associated
with these models to describe bremmstrahlung emission and the SZE.
\S~\ref{sec:sample} describes our expanded cluster sample, and
\S~\ref{sec:dataanalysis} the details of the x-ray data analysis, our
Monte Carlo error analysis, and the error budget.  In this section and
the following one, we pay special attention to the effects of central
cooling flow emission on our profile models.  We present our results
in \S~\ref{sec:results}, including mass models and a quantitative
assessment of the uncertainties in the SZE predictions, as well as
total masses and baryon mass fractions derived under the assumption of
hydrostatic equilibrium.  We discuss our results and conclude in
\S~\ref{sec:conclusion}.  Throughout we use $H_o = 100 \, h^{-1} {\rm
km/sec/Mpc}$ and $q_o = \frac{1}{2}$ unless otherwise specified; we
will comment on the impact that assuming other cosmologies has on our
results in \S~\ref{sec:conclusion}.

We will report an improved measurement of $H_o$ using the observations
of Myers et al. and \citet{Herbig_et_al_1995} plus recent observations
of Abell 399 in a second paper (paper II).  This measurement relies
upon the density models we present in this paper, as well as improved
electron temperatures from the literature.  The consequences of this
x-ray analysis in the context of our measurement of $H_o$ will also be
discussed in paper II.

\section{Models of the Intra-Cluster Medium}
\label{sec:icmmodels}

One model for the cluster gas which has enjoyed great phenomenological success is the {\it isothermal beta model} \citep{Cavaliere_and_Fusco_Fermiani_1976}.
In this case, the gas is assumed to be isothermal, and the electrons distributed according to 
\begin{equation}
n_e(r) = n_{eo} \, \left( 1 + r^2/R_0^2\right)^{-3 \beta/2}.
\label{eq:betamod}
\end{equation}
Here $n_{eo}$ is the central electron number density, $r$ is the
spherical metric radius, and $R_0$ is a characteristic scale.  The
properties of this model are well-known and extensively tabulated in
the literature.  We have assumed that the ICM is spherically
symmetric.  While this is not generally the case for individual
clusters, it should be a good description on average for an
orientation-unbiased sample (\S~\ref{sec:sample}).

One generalization of the isothermal beta model which has some
support both observationally \citep[e.g.][]{Hughes_et_al_1988b} and
theoretically \citep[e.g.][]{Navarro_Frenk_and_White_1997} is the
hybrid isothermal/adiabatic model.  In paper II we will study the
effects of these models on the SZE, but since the ROSAT data are not
sensitive to temperature gradients in the outer parts of the cluster,
we will assume an isothermal ICM for the remainder of this
analysis.  Some authors
\citep[e.g.][]{Gunn_and_Thomas_1996,Holzapfel_et_al_1997,Daisuke_Sulkanen_and_Evrard_1999}
have studied the possibility of a multi-phase ICM.  In
\S~\ref{sec:spectralanalysis} we will present some preliminary
indications of cool ICM phases we have found in our analysis.  It is,
however, difficult to disentangle the effects of these phases from
ROSAT calibration uncertainties.  We will assess the impact of the
observed effect in \S~\ref{sec:spectralanalysis}, but until better
data from future missions is capable of accurately constraining ICM
phase models, we will adhere to the single-phase model.

\subsection{Thermal Bremmstrahlung}

The bolometric luminosity due to bremsstrahlung emission 
from an ionized thermal plasma of electrons and protons is \citep{Rybicki_and_Lightman_1979}
\begin{equation}
L = W \times \int_{\it ICM} n_e n_p T_e^{1/2} \overline{g}(T_e) \, dV , 
\label{eq:bremlum}
\end{equation}
where 
\begin{equation}
W = \left( \frac{2 \pi k_B}{3 m_e} \right)^{1/2} \times \frac{2^5 \pi e^6}{3 h m_e c^3}  , 
\end{equation}
$n_e$ is the electron number density, $n_p$ is the proton number density, $k_B$ is Boltzmann's
constant, $T_e$ is the local temperature of the plasma, and $\overline{g}(T_e)$ is the
thermally averaged Gaunt factor.   For a plasma with the cosmic helium mass fraction $Y = 0.24$ and metal
abundances of 30\% solar, $n_p/n_e = 0.862$;  this is the value we adopt.  This corresponds
to a baryonic mass per electron of 
$\mu_e = 1.146$ and an overall mean molecular mass (in units of the
proton mass) $\mu = 0.592$; these calculations assume the solar abundances of \citet{Anders_and_Grevesse_1989}.

For an isothermal plasma, the bolometric flux observed at a redshift $z$ is
\begin{equation}
S = \frac{W \overline{g}(T_e) \, T_e^{1/2}}{4 \pi D_L^2(z)} \, \times \int n_e n_p dV.
\end{equation}
where 
$D_L$ is the luminosity distance \citep{Weinberg_1972}
\begin{equation}
\label{eq:luminosity_distance}
D_L = 6000 \, h^{-1} \, \left[ (1+z) - \sqrt{1+z} \right] {\rm Mpc}
\end{equation}
(for $q_o = 1/2$).  The integral on the right hand side is the Emission Measure (EM):
\begin{equation}
EM =  \int n_e n_p dV.
\end{equation}
For the beta model, the EM over all space reduces to
\begin{equation}
EM = \frac{n_p}{n_e} \,\pi^{\frac{3}{2}} r_o^3 n_{eo}^2 \frac{\Gamma(3 \beta - 3/2)}{
\Gamma(3 \beta)}
\end{equation}
If we know the form of the electron density profile $(\theta_o,\beta)$, the cluster temperature $T_e$,
the redshift, and the bolometric flux, we may determine the normalization of the density profile:
\begin{equation}
\label{eq:theoretical_density}
n_{eo} = \sqrt{ \frac{n_e}{n_p} \, \frac{4}{\pi^{1/2}} \frac{\Gamma (3 \beta)}{\Gamma (3 \beta - \frac{3}{2})} \frac{1}{\theta_o^3}  \frac{(1+z)^6}{D_L} \frac{S}{W \overline{g}(T_e)\, T_e^{1/2}}}.
\end{equation}
Here we have used the fact that $D_L = D_A \times (1+z)^2$.

In practice the x-ray spectrum is observed over a finite bandpass with
a finite aperture, and is modified by photoelectric absorption due to
the intervening intra-Galactic medium.  We account for this by
modeling the spectrum with {\tt XSPEC}, NASA GSFC's standard x-ray
spectral analysis program.  {\tt XSPEC} reports the normalization of
the spectrum (a quantity analogous to $S$ in the foregoing discussion)
through the parameter $K$, defined as
\begin{equation}
\label{eq:raymond_k}
K = \frac{10^{-14}}{4 \pi D_L^2} \, \int \, n_e \, n_p \, dV.
\end{equation}
Here all distances are in cm.  The EM in the annulus between $\theta_1$ and $\theta_2$ is
\begin{equation}
\scriptsize
\label{eq:aperture_integral}
EM(\theta_1,\theta_2) =  \frac{n_p}{n_e} n_{eo}^2 D_A^3 \theta_o^3 \sqrt{\pi}
 \frac{\Gamma(3\beta-1/2)}{ \Gamma(3\beta)} \times \frac{2 \pi}{3(2 \beta -1)} \, (C(\theta_1) - C(\theta_2) )
\end{equation}
where
\begin{equation}
\label{eq:c_factor}
C(\theta) = \left[ 1 + \left(\frac{\theta}{\theta_o}\right)^2 \right]^{-3 \beta + 3/2}.
\end{equation}
The central density is then
\begin{eqnarray}
\label{eq:density}
\scriptsize
n_{eo} & = & 4.160 \times 10^{-3} \, {\rm cm^{-3}} \, h^{1/2} \, \times \\ & & \left[ \frac{n_e/n_p}{1.16} \, (\beta - 1/2) \, 
\frac{\rm arcmin^3}{\theta_o^{3}} \,  \frac{(1+z)^6}{1+z -  \sqrt{1+z}} \frac{\Gamma (3 \beta)}{\Gamma (3 \beta - 1/2)} 
\frac{K/10^{-2}}{C(\theta_1) - C(\theta_2)} \right]^{\frac{1}{2}} \nonumber.
\end{eqnarray}
Note that Eq.~\ref{eq:luminosity_distance} used in deriving this equation assumed
$q_o=\frac{1}{2}$; for $z = 0.1$ this causes at most a $\pm1\%$ error in $n_{eo}$ for $\Delta q_o = \pm \frac{1}{2}$.
This is the expression we use to determine the normalization of the cluster density profiles.  

\subsection{The Sunyaev-Zeldovich Effect}

\citet{Sunyaev_and_Zeldovich_1980} show that the fractional change in intensity due to
the inverse Compton scattering of CMB photons by a thermal plasma with a velocity $v_r$ along
the line of sight is 
\begin{equation}
\label{eq:sze_full_form}
\frac{\Delta I_{\nu}}{I_{\nu}}  =  \tau \frac{ x e^x}{e^x - 1} \times \left( \frac{k T_e}{m_e c^2} f(x) +
 \frac{v_r}{c} + \Theta(T_e,v_r) \right) , 
\end{equation}
where $T_e$ is the temperature of the plasma, $\tau$ is the optical depth for inverse Compton scattering
\begin{equation}
\label{eq:tau}
\tau = \sigma_T \, \int \, n_e(z) \, dz,
\end{equation}
 $x$ is the dimensionless frequency 
\begin{equation}
x = \frac{h \nu}{k T_{cmb}},
\end{equation}
and 
\begin{equation}
f(x) = x \coth (x/2) - 4.
\end{equation}
$\Theta(T_e,v_r)$ is a relativistic correction \citep{Rephaeli_1995,
Challinor_and_Lasenby_1998,Sazonov_and_Sunyaev_1998}, which generally
has a magnitude of a few percent that of the leading term for $T_e
\lesssim 10 \keV$.  In Eq.~\ref{eq:tau}, $\sigma_T$ is the Thompson
scattering cross section ($6.65 \times 10^{-25} {\rm cm^2}$) and the
integral is along the line of sight.  Since the ROSAT PSPC data are
not sensitive to temperature gradients, they are capable of directly
constraining the cluster density profiles.  For this reason we express
our SZE analysis in terms of the inverse Compton optical depth.  Note
that this also renders the results we present here independent of the
relativistic correction $\Theta(T_e,v_r)$, although this will have to
be taken into account when relating the $\tau$ values we present 
to observed intensity decrements.

A radio telescope will measure the intensity decrement convolved with
the instrument beam pattern.  It is then convenient to consider the
quantity
\begin{equation}
\tau_{beam} = \frac{1}{\Omega_{Beam}}  \, \int \, d\Omega \, \tau(\hat{\Omega}) \, R_N(\hat{\Omega})
\end{equation}
where $R_N(\hat{\Omega})$ is the telescope beam normalized to unity at the maximum.  $R_N$ is typically
well described by a Gaussian:
\begin{equation}
R_N(\hat{\Omega}) = e^{-\theta^2/2 \sigma^2}.
\end{equation}
For the OVRO 5.5-meter telescope we have $\sigma_{\scriptsize 5.5-m} =
3'.12 \pm 0.11$ \citep{Leitch_thesis}, and for ACBAR,
$\sigma_{\scriptsize ACBAR} = 1'.7$.  The 5.5-meter telescope also
switches by $22'.16$ in azimuth to remove atmosphere and ground
emission.  We then define
\begin{equation}
R_{sw}(\hat{\Omega}) = R_N(\hat{\Omega}) - R_N(\hat{\Omega} - \delta\hat{\Omega} )
\end{equation}
where $\delta \hat{\Omega}$ is the switching vector which, at a given Hour Angle, corresponding to a $22'.16$ offset
in azimuth.  Since our analysis assumes spherical symmetry for the clusters, we will always compute this quantity
at transit.  The equivalent quantity to $\tau_{beam}$ then is
\begin{equation}
\tau_{sw} = \frac{1}{\Omega_{Beam}}  \, \int \, d\Omega \, \tau(\hat{\Omega}) \, R_{sw}(\hat{\Omega}).
\end{equation}

In terms of these objects, the equivalent temperature decrement which a single-dish telescope will see
for an isothermal cluster is 
\begin{equation}
\Delta T_{eq} = T_{cmb} \, \frac{x^2 \, e^x}{(e^x-1)^2} \, f(x) \, \frac{k T_e}{m_e c^2} \, \tau_{sw}.
\end{equation}
For more details on this formalism, see \citet{Myers_et_al_1997} and
\citet{Mason_thesis}.

\section{The Sample}
\label{sec:sample}

Due to the assumption of spherical symmetry necessary to employ the
Sun\-yaev-Zeldovich effect as a distance measure, it is vital to
measure the SZE in an unbiased sample. An x-ray selected, x-ray flux
limited sample satisfies this criterion.  The first steps in this
direction were taken by \citet{Myers_et_al_1997} who used the sample
of \citet{Edge_et_al_1990} to define an x-ray flux limited sample of
11 clusters with $z < 0.1$ , $f_x > 3.11 \times 10^{-11} \ergcms \, $
(2-10 keV), galactic latitudes $\mid b \mid > 20^{\circ}$, and
declinations $\delta > -23^{\circ}$.  We improve upon this by defining
a larger sample.  This has two advantages.  First, it will allow the
(random) noise caused by departures from spherical symmetry to be
reduced in the sample average.  Second, it will allow many possible
systematic affects to be identified by dividing the sample into
several sub-samples by, {\it e.g.}, morphology, optical richness, or
x-ray spectral characteristics.

As our parent sample we choose the XBAC catalog of
\citet{Ebeling_et_al_1996}.  Ebeling et al. have analyzed the data of
the Rosat All Sky Survey (RASS) to obtain a sample of 242 Abell
clusters complete at the $95\%$ level down to a flux limit of $5.0
\times 10^{-12} \ergcms$ in the 0.1--2.4 keV band.  If we
conservatively adopt a flux cutoff of twice this ($1.0 \times 10^{-11}
\ergcms$) and impose the luminosity cutoff ($D_L(z=0.1) = 291.52 \,
h^{-1} \, {\rm Mpc}$ for $q=1/2$) corresponding to a volume complete
sample out to $z=0.1$, we find 31 clusters in the XBAC catalog that
meet these criteria.  These are listed in
Table~\ref{tbl:clustersample}.

Since the XBAC catalog ultimately uses the list of Abell clusters as a
parent sample, we must employ some caution in assessing the
completeness of our sample.  By comparing the overlap regions of the
Abell \citep{Abell_1958} and ACO \citep{ACO_1989} catalogs,
\citet{Scaramella_et_al_1991} find that Abell missed 29\% of the
Richness class 1 or greater clusters which were in ACO; most of these
are $R=1$ objects.  These results are consistent with the results of
\citet{van_Haarlem_et_al_1997}, who find that $30\%$ of $R>=1$ objects
are missed in Monte Carlo simulations with an optical selection
algorithm designed to mimic that of Abell and ACO.  In most cases,
these clusters are missed due to fluctuations in the background galaxy
counts, resulting in the misclassification of actual clusters as
poorer objects which are then not included in the catalog; since this
misclassification preferentially affects the poorer clusters, most of
the missed clusters are again likely to be Richness class 1 objects.
Our stringent luminosity criterion strongly selects against these
poorer objects: only $23\%$ of our 31 cluster sample are $R <=1$,
whereas $67\%$ of the 242 clusters in the XBAC catalog have $R <=1$.
Van Haarlem et al. find that a luminosity cutoff comparable to the one
we have imposed, applied after the optical selection, results in a
catalog which is free from false detections (although this does not
directly address the issue of orientation bias).

Based on these considerations, we estimate our sample to be 90\%
complete for $L_o = 1.13 \times 10^{44} h^{-2} \ergs$, $f_o = 1.0
\times 10^{-11} \ergcms$.  This corresponds to 3 missed clusters at
most.  This level of incompleteness will not significantly affect
$H_o$ measurements derived from the sample as a whole.  Near-future,
fully x-ray selected surveys such as the ESO REFLEX survey will
provide an important cross-check on the completeness of this sample.

%
%
\renewcommand{\baselinestretch}{1}\small\normalsize 
\begin{table}
\begin{center}
\footnotesize
\begin{tabular}{r c c c r r}
\hline\hline
 Source&  RA         & Dec               &$F_{x}$& $L_{x}$   \\ \hline
 A2142 &  15:58:22.1 & +27:13:58.8     & 61.4  & 20.74     \\ 
 A2029 &  15:10:55.0 & +05:43:12.0     & 61.6  &  15.35    \\ 
 A478  &  04:13:26.2 & +10:27:57.6     & 39.1  &  12.95    \\ 
 A1795 &  13:48:52.3 & +26:35:52.8     & 67.2  &  11.12    \\ 
 A401  &  02:58:56.9 & +13:34:22.8     & 42.6  &  9.88     \\ 
 A2244 &  17:02:40.1 & +34:03:46.8     & 22.8  &   9.09    \\ 
 A3667 &  20:12:23.5 & $-$56:48:46.8   & 73.1  &   8.76    \\ 
 A85   &  00:41:48.7 & $-$09:19:04.8   & 72.3  &    8.38   \\ 
 A1651 &  12:59:24.0 & $-$04:11:20.4   & 27.1  &  8.25     \\ 
 A754  &  09:09:01.4 & $-$09:39:18.0   & 64.1  &   8.01    \\ 
 A2597 &  23:25:16.6 & $-$12:07:26.4   & 25.9  &   7.97    \\ 
 A1650 &  12:58:41.8 & $-$01:45:21.6   & 25.6  &   7.81    \\ 
 A3827 &  22:01:56.6 & $-$59:57:14.4   & 18.7  &   7.78    \\ 
 A3112 &  03:17:56.4 & $-$44:14:16.8   & 36.4  &   7.70    \\ 
 A3571 &  13:47:28.1 & $-$32:51:14.4   & 109.5 &  7.36     \\ 
 A1656 &  12:59:31.9 & +27:54:10.8     & 316.5 & 7.21      \\ 
 A2256 &  17:04:02.4 & +78:37:55.2     & 49.0  &  7.05     \\ 
 A2384 &  21:52:16.6 & $-$19:36:00.0   & 18.2  &   6.82    \\ 
 A780  &  09:18:06.7 & $-$12:05:56.4   & 48.4  &    6.63   \\ 
 A399  &  02:57:49.7 & +13:03:10.8     & 29.0  &  6.45     \\ 
 A3558 &  13:27:57.8 & $-$31:29:16.8   & 64.6  &   6.27    \\ 
 A3266 &  04:31:25.4 & $-$61:25:01.2   & 48.5  &   6.15    \\ 
 A4010 &  23:31:14.2 & $-$36:30:07.2   & 14.1  &   5.55    \\ 
 A3921 &  22:49:59.8 & $-$64:25:51.6   & 14.0  &   5.40    \\ 
 A3158 &  03:42:43.9 & $-$53:38:27.6   & 35.7  &   5.31    \\ 
 A2426 &  22:14:32.4 & $-$10:21:54.0   & 12.2  &   5.10    \\ 
 A3695 &  20:34:46.6 & $-$35:49:48.0   & 15.1  &   5.07    \\ 
 A2065 &  15:22:26.9 & +27:42:39.6     & 22.3  &   4.95    \\ 
 A2255 &  17:12:45.1 & +64:03:43.2     & 17.2  &   4.79    \\ 
 A566  &  07:04:22.3 & +63:16:30.0     & 11.3  &    4.62   \\ 
 A3911 &  22:46:20.9 & $-$52:43:30.0   & 11.8  &   4.61    \\ 
\hline 
\end{tabular} 
\caption{SZE cluster XBAC subsample.  Columns are:    Right Ascension and Declination (J2000),
$F_{x}$ , $ 10^{-12} \ergcms $ , 
$ 0.1-2.4 \keV $ ; $L_{x}$  , $ 10^{44} \ergsh2 $.  Fluxes and luminosities are as reported by
Ebeling et al. and assume $h=0.5$.}
\label{tbl:clustersample}
\end{center}
\end{table}
\renewcommand{\baselinestretch}{1.6}\small\normalsize \phantom{z}

%
%
\renewcommand{\baselinestretch}{1}\small\normalsize 
\begin{table}
\begin{center}
\begin{tabular}{l l l c}
\hline\hline
Cluster & $T_e \,\,$    & $N_H$ & $z$ \\
        &  keV     & $10^{20} {\rm cm^{-2}}$ & \\
\hline
A85   & $6.9\pm 0.4$     &  3.44  & 0.0518  \\
A399  & $7.0\pm 0.4$     &  10.9  & 0.0715  \\
A401  & $8.0\pm 0.4$     &  10.5  & 0.0748  \\
A478  & $8.4^{+0.8}_{-1.4}$& $21.1^{\ast}$ & 0.0900  \\
A754  & $9.5^{+0.7}_{-0.4}$& 4.36 & 0.0528  \\
A780  & $4.3\pm 0.4$     & 4.94   & 0.0522  \\
A1651 & $6.1\pm 0.4$     & 1.81  & 0.0825  \\
A1656 & $9.1\pm0.7 ^{A}$ & 0.92    & 0.0232  \\
A1795 & $7.8\pm 1.0$     & 1.19  & 0.0616  \\
A2029 & $9.1\pm 1.0$     & 3.06  & 0.0767  \\
A2142 & $9.7^{+1.5}_{-1.1}$& 4.20  & 0.0899  \\
A2244 & $7.1^{+5.0 \, \, B}_{-2.2}$& 2.13  & $0.0980$ \\
A2255 & $7.3^{+3.3 \, \, B}_{-1.6}$& 2.59 & $0.0800$\\
A2256 & $6.6 \pm 0.4$   & 4.10   & 0.0601 \\
A2597 & $4.4^{+0.4}_{-0.7}$ & 2.49  & 0.0852 \\
A3112 & $5.3^{+0.7}_{-1.0}$ & 2.60  & $0.0703^{C}$  \\
A3158 & $5.5 \pm 0.6^{B}$& 1.35  & $0.0590^{C}$  \\
A3266 & $8.0\pm 0.5$      &  1.60 &  $0.0594^{C}$ \\
A3558 & $5.5\pm 0.4$      & 3.88  &  $0.0482^{C}$ \\
A3571 & $6.9 \pm 0.2$     & 3.70  &  $0.0397^{D}$ \\
A3667 & $7.0 \pm 0.6$     & 4.76  &  $0.0552^{E}$ \\
A3921 & $6.6 \pm 1.6^{F}$ & 2.95      &    $0.0960^{G}$ \\
\hline
\multicolumn{4}{l}{\footnotesize $\ast$ ROSAT value; Stark et al. gives $14.8$} \\
\multicolumn{4}{l}{\footnotesize$^A$ \citet{Hughes_et_al_1988a}} \\
\multicolumn{4}{l}{\footnotesize$^B$ \citet{David_et_al_1993} } \\
\multicolumn{4}{l}{\footnotesize$^C$ \citet{ACO_1989} } \\
\multicolumn{4}{l}{\footnotesize$^D $ \citet{a3571_redshift} } \\
\multicolumn{4}{l}{\footnotesize$^E$ \citet{a3667_redshift} } \\
\multicolumn{4}{l}{\footnotesize$^F$ \citet{Ebeling_et_al_1996} with our estimated error. } \\
\multicolumn{3}{l}{\footnotesize$^G$ \citet{a3921_redshift} } \\
\end{tabular}
\caption{Electron temperatures
are from \citet{Markevitch_et_al_1998}  except as noted; redshifts are from \citet{Struble_and_Rood_1991_redshifts}
except as noted.  \label{tbl:redshift_temp}  All errors are 90\% ($\sim 1.65-\sigma$) confidence limits.}
\end{center}
\end{table}
\renewcommand{\baselinestretch}{1.6}\small\normalsize \phantom{z}

In Table~\ref{tbl:redshift_temp} we summarize redshift and temperature
data from the literature on the 22 clusters with public ROSAT data in
our sample.  Where available, we use the \citet{Markevitch_et_al_1998}
ASCA temperatures which properly account for cooling flow
contamination of the x-ray spectrum.  We also list the Hydrogen column
densities employed in the spectral analysis
(\S~\ref{sec:spectralanalysis}).  With the exception of A478 we use
the Galactic neutral Hydrogen values of \citet{Stark_et_al_1992} as
interpolated onto the cluster coordinates by {\tt PIMMS/COLDEN}.  For
A478 we use the column density resulting from our x-ray spectral
analysis; this is discussed further in \S~\ref{sec:spectralanalysis}.
While the 22 clusters in Table~\ref{tbl:redshift_temp} are not
strictly speaking a flux-limited sample, we note that ROSAT targets
are preferentially high-luminosity clusters (as opposed to
morphologically selected clusters) and thus the 22 cluster sub-sample
is likely to be free from orientation effects also.  Three of the 9
missed clusters have non-public ROSAT observations (A4010, A2426, and
A3695) and one (A2384) has HRI observations only; these four
observations plus future XMM observations for which we will propose
will help fill out the sample.

\clearpage
\section{Data Analysis}
\label{sec:dataanalysis}

We searched the public ROSAT data archive for observations of the 31
clusters in Table~\ref{tbl:clustersample} and found a total of 44
observations on 22 clusters.  These data form the basis for our
analysis; the data sets are listed in Table~\ref{tbl:observations},
along with the ROSAT sequence ID, exposure time, approximate pointing
offset from the cluster center, and date of the ROSAT observation.  We
assign an observation tag (e.g., A754d) to each observation for
convenience.  After the date we indicate the detector used for each
observation: PSPC B (B), PSPC B/High gain (BH), or PSPC C (C). We will
use the redundancy of the data to search for possible ROSAT
calibration errors.

For each observation of a  cluster with a single instrument (PSPC B and BH--- there
were no multiple observations with the PSPC C),
all individual pointings were mosaicked using the ESAS software described below
to form a single count rate image for the cluster.  These data form the basis for our
highest signal-to-noise analyses where available.  These mosaicked observations
are referred to by the tags BM and  BHM; a catalog of the mosaics is presented in
Table~\ref{tbl:mosaic_cat}.  After this step, the mosaics are analyzed in the
same fashion as the single-pointing observations.

In order to clean the data and correct for telescope vignetting, as
well as for the mosaicking, we used the Extended Source Analysis
Software (ESAS); see 
\citet{Snowden_et_al_1994}  for more details on
this package.  Table~\ref{tbl:snowden_bands} shows the energy channel definitions
employed by this software.  While in principle the
cluster luminosity may be inferred by observations over a single band
(e.g., by co-adding R4--R7), we chose to analyze each band separately.
This allows us to check the suitability of our x-ray models more
carefully.  All data with Master Veto count
rates greater than 170 counts per second were rejected.  The end
product of the ESAS analysis are 512 x 512 pixel maps (each pixel
$14''.947$ on a side) of the cleaned, vignetting corrected count rate,
the exposure, and the raw counts.  ESAS also provides a map of the
background model used in cleaning the data.  These products are
provided for each of the seven analysis channels, and also for the
co-added R4--R7 channel which we use for the profile analysis.

Point source masks were derived using ESAS's {\tt DETECT} algorithm,
which employs a variable detection aperture to account for the change
in the ROSAT point-spread function across the detector face.  We
masked all sources detected with greater than 99\% confidence in the
0.5 -- 2.0 keV band having total count rates greater than $\,
5\times 10^{-3} \,$ counts per second.  This is potentially important,
since the variability of the point spread function and the detector
vignetting are likely to bias point source masks created ``by eye''.
Using the exposure maps calculated by ESAS, all regions with exposures
less than $30\%$ of the maximum are also masked; this removes data for
which the vignetting correction is more than a factor of $\sim 3$.
Typically $\sim 10\%$ of the pixels are masked by this step, primarily
those shadowed by the PSPC window support structure.  These masks were
employed to exclude data in the subsequent profile and spectral
analysis.

%
%
\renewcommand{\baselinestretch}{1}\small\normalsize 
\begin{table}
\begin{center}
\scriptsize
\begin{tabular}{ l l l l l l }
\hline\hline
Cluster & Sequence ID & Observation & Exposure & Offset & Observation \\ 
        &             &    Tag      &   (sec)  & (arcmin)&  Date        \\ \hline
{\bf A85}& rp800250N00 & a85a      & 10238          & 4.74  & 01jul92 (B)\\ \hline
 	& rp800174A00 & a85b      & 2187           & 5.20  & 20dec91 (B)\\ \hline
 	& rp800174A01 & a85c      & 3458  & 5.20 & 11jun92 (B)\\ \hline
{\bf A401} & rp800182N00 & a401a & 6735   & 0.22 & 23jan92 (B)\\ \hline
{\em (incl. A399)} & rp800235N00 & a401b & 7457 & 0.22 & 30jul92 (B)\\ \hline
{\bf A478}    & rp800193N00 & a478a     & 21969          & 0.4 & 31aug91 (BH)\\ \hline
{\bf a754}  & rp800550N00 & a754a & 8156 & 12.13 & 06nov93 (B)\\ \hline
 	& rp600451N00  & a754b    & 13495           & 22.34 & 03nov92 (B)\\ \hline
 	& rp800160N00  & a754c    & 2266           & 12.13  & 19nov91  (B)\\ \hline
 	& rp800232N00  & a754d    & 6358           & 12.13  & 10nov92  (B)\\ \hline
{\bf A780}    & rp800318n00 & a780a     & 18398          & 12.00          & 08nov92  (B)\\ \hline
{\bf A1651} & wp800353  & a1651a  & 7435           & 3.41   & 18jul92  (B)\\ \hline
{\bf A1656}   & rp800009n00 & a1656a    & 20345          & 33.04          & 16jun91  (BH)\\ \hline
              & rp800006n00 & a1656b    & 21545          & 9.98           & 16jun91  (BH)\\ \hline
              & rp800005n00 & a1656c    & 21140          & 2.23           & 17jun91  (BH)\\ \hline
              & rp800013n00 & a1656d    & 21428          & 15.85          & 18jun91  (BH)\\ \hline
{\bf A1795} & rp700284N00 & a1795a & 2025          & 13.84  & 30jun91  (BH)\\ \hline
 	& rp700145A01  & a1795b   & 1909           & 13.84  & 06jan92 (B)\\ \hline
 	& rp700145A00  & a1795c   & 18205          & 13.84  & 01jul91 (BH)\\ \hline
 	& rp800105N00  & a1795d   & 36273          & 0.58   & 04jan92 (B) \\ \hline
 	& rp80055N00   & a1795e   & 25803          & 1.67   & 09jul91  (BH)\\ \hline
{\bf A2029} & rp800161N00 & a2029a & 3151          & 1.12   & 24jan92  (B)\\ \hline
 	& rp800249N00  & a2029b   & 12542          & 0.32   & 10aug92  (B)\\ \hline
{\bf A2142}   & rp800415N00 & a2142a    & 19208          & 15.78 &  21aug92  (B)\\ \hline
 	& rp150084N00 & a2142c    & 7734           & 0.46  & 20jul90  (C)\\ \hline
  	& wp800096    & a2142d    & 6192           & 0.46  & 25aug92  (B)\\ \hline
 	& rp800551N00 & a2142e    & 6090           & 0.46  & 23jul93  (B)\\ \hline
	& rp800233N00 & a2142f    & 4939           & 0.46  & 26aug92  (B)\\ \hline
{\bf A2244}   & rp800265N00 & a2244a    & 2963           & 1.82           & 21sep92  (B)\\ \hline
{\bf A2255}   & rp800512n00 & a2255a    & 14555          & 1.64           & 24aug93  (B)\\ \hline
{\bf A2256}& rp100110N00 & a2256a    & 17032          & 0.18  & 17jun90  (C)\\ \hline
 	& rp800163N00 & a2256b    & 10681          & 14.51 & 25nov91  (B)\\ \hline
 	& rp800340N00 & a2256e    & 9422           & 22.59 & 25jul92  (B)\\ \hline
        & rp800341N00 & a2256f    & 10473          & 16.54 & 23jul92  (B)\\ \hline
	& rp800162A00 & a2256g    & 4246           & 13.35 & 15oct91  (B)\\ \hline
	& rp800162A01 & a2256h    & 4747           & 13.35 & 15mar92  (B)\\ \hline
 	& rp800339N00 & a2256i    & 4978           & 13.35 & 22jul92  (B)\\ \hline
{\bf A2597}   & rp80012N00  & a2597a    & 7163           & 1.12           & 27nov91  (B)\\ \hline
{\bf A3112}  & rp800302N00 & a3112a & 7598 & 4.50 & 17dec92  (B)\\ \hline
{\bf A3158}   & rp800310n00 & a3158a    & 3020           & 19.08          & 26aug92  (B)\\ \hline
{\bf A3266 } & rp800552N00 & a3266a & 13547 & 2.14 & 19aug93  (B)\\ \hline
{\bf A3558}   & rp800076n00 & a3558a    & 29490          & 1.48           & 17jul91  (BH)\\ \hline
{\bf A3571}   & rp800287n00 & a3571a    & 6062           & 12.08          & 12aug92  (B)\\\hline
{\bf A3667}   & rp800234n00 & a3667a    & 12550          & 0.26           & 09oct92 (B)\\ \hline
{\bf A3921}   & rp800378n00 & a3921c    & 11997          & 1.19           & 15nov92  (B)\\ \hline
\end{tabular}
\caption{ROSAT PSPC observations of clusters in our sample.  Shown are the cluster name, the ROSAT Sequence ID, the observation tag we assign to the observations, the exposure time in seconds, the offset of the ROSAT pointing center from
the cluster center, and the date of the ROSAT observation.}
\label{tbl:observations}
\end{center}
\end{table}
\renewcommand{\baselinestretch}{1.6}\small\normalsize \phantom{z}

\begin{table}[t!]
\begin{center}
\begin{tabular}{l l }
\hline\hline
Mosaic & Pointings \\ \hline
A2142BM & A2142A,E,F \\
A2256BM & A2256B,E--I \\
A85BM         &  A85A--C\\
A401BM        &  A401A,B\\
A754BM        &  A754A--D\\
A2029BM      &  A2029A,B\\
A1795BHM    & A1795A,C,E \\
\hline
\end{tabular}
\end{center}
\caption{A catalog of the mosaics used in the spectral analysis.  We show
the mosaic tag that we use for reference, and the individual observations
composing each mosaic.}
\label{tbl:mosaic_cat}
\end{table}

\begin{table}[b!]
\begin{center}
\begin{tabular}{l l l}
\hline\hline
Band  &  Energy & Centroid \\
Name  &  (keV)  &   (keV) \\ \hline
R1 & 0.11 -- 0.284 & 0.197 \\
R2 & 0.14 -- 0.284 & 0.212 \\
R3 & 0.20 -- 0.83\phantom0 & \phantom{1l}--- \\
R4 & 0.44 -- 1.01\phantom0 & 0.725 \\
R5 & 0.56 -- 1.21\phantom0 & 0.850 \\
R6 & 0.73 -- 1.56\phantom0 & 1.140 \\
R7 & 1.05 -- 2.04\phantom0 & 1.540 \\
\hline
\end{tabular}
\end{center}
\caption{Snowden PSPC energy band defintions, along with their approximate centroids.  Since band 3
straddles an absorption edge due to carbon in the PSPC window, it is not significant to
designate a centroid for this band.  This band is not used in any subsequent analysis.}
\label{tbl:snowden_bands}
\end{table}

\clearpage

\clearpage
\subsection{Profile Analysis}
\label{sec:profileanalysis}

We used the composite R4--R7 count rate image to determine the spatial
profile of each cluster.  Where mosaicked observations were possible,
we used these.  The centroid of each image was computed in a circle
centered on the visual peak of emission\footnote{The only exception to
this was A754, where we varied the center used for the azimuthal
average until a good beta model fit could be obtained.  The
resulting centroid is $\alpha=137^{\circ}.314,\delta=-9^{\circ}.674
\,$ (J2000).  The uncertainty induced by this procedure will be
accounted for in our Monte Carlo simulations
(\S~\ref{sec:montecarlosimulations})}, and an azimuthal profile was
constructed with 150 bins out to a radius equivalent to $1.5 h^{-1}
\,$ Mpc or $50'$, whichever was less.  We find that our profile
results are not sensitive to the binning.  Error bars were assigned by
performing an identical procedure on the co-added R4--R7 raw count
images, and assigning the fractional Poisson error for each bin in the
raw count profile (based on the total number of counts in that bin) to
the corresponding bin in the count-rate profile.

The resulting profile was fit to a beta model surface brightness profile:
\begin{equation}
I(\theta) = I_o \times \left( 1 + \frac{\theta^2}{\theta_0^2}\right)^{-3 \beta + \frac{1}{2}}.
\end{equation}
For cooling flow clusters in which a good fit could not be obtained
for a beta model, we added a Gaussian component of emission to the
profile (see below).  In these fits, $\theta_0$, $\beta$, $I_o$, and
the central intensity and width of the Gaussian were all free
parameters.  A free constant background term was also included. The
best-fit parameter values were obtained using a standard non-linear
least-squares code \citep{Numerical_Recipes}, and the uncertainties in
these parameters were determined from our Monte Carlo analysis
(\S~\ref{sec:montecarlosimulations}).  In order to protect the
assumption of Gaussianity implicitly assumed by the least-squares
method, any bins in the profile with fewer than 16 counts total were
rejected. This precaution was only necessary for some of the shortest
exposures (e.g., A2244a); more typically, all of the bins in the
profile had $\gtrsim 50$ counts.

In \S~\ref{sec:montecarlosimulations} we will describe the Monte Carlo
simulations which we use to establish the confidence intervals for the
profile parameters.  Note, however, that the Monte Carlo technique
assumes at the outset some parametric model for the cluster profile,
which is then fit to multiple simulated data sets.  If the chosen
profile is simply not appropriate, that fact may not be reflected in
the resultant confidence intervals. 
This issue is especially important for the cooling flow clusters: a
single beta model does not typically fit the azimuthal profile of
these clusters over the full range of interesting radii, and the fits
tend to be strongly driven by the high signal-to-noise data in the
cluster core.  The literature shows that investigators have developed
a number of strategies to model these clusters.  For example,
\citet{Briel_and_Henry_1996_1795_tmap} and
\citet{Henry_and_Briel_1996}, in modeling Abell 1795 and 2142,
exclude the inner $3'- 5'$ of the profile and fit the outer part to a
standard beta model.  \citet{Mohr_et_al_1999} model the cooling
flow clusters in their sample as the sum of two beta model profiles
in emission, and numerically solve for the underlying density profile.

For the cooling flow clusters in our sample we chose to construct two
models for each cluster using methods similar to each of these
strategies.  This analysis allows us to quantitatively evaluate the
impact of our cooling-flow modeling strategy on the scientific
results we present.  The first method we use is to fit the projected
emission profile to a beta model plus a Gaussian component (to
represent the central excess of emission over the beta model).
Since modeling the central emission is highly uncertain because of
the complicated physics in this region, we ignore the fitted excess in
constructing density profiles for the subsequent analysis\footnote{A
comparison with Mohr et al.  (to be provided in
\S~\ref{sec:conclusion}) will show up any systematic errors in the
baryonic masses due to this procedure, as these authors retain the
central excess and solve for the underlying 3-D profile assuming a
single-phase ICM.}.  Our second strategy (similar to that of Briel and
Henry) is to excise data out to a radius of $150 \, h^{-1}$ kpc in the
image and fit a beta model to the remaining data.  Note that this
method provides a less core-weighted measurement than the former
method, and since the SZE tends to be dominated by the emission from
the outer parts of the cluster, this is an important diagnostic.  The
same considerations pertain for the total mass measurement, which
depends only upon the derivative of the density profile at some
(typically large) radius.

The former (beta plus Gaussian component) models we will refer to as
the ``primary'' models for the cooling flow clusters.  In the final
analysis all of our baryon models (and therefore baryonic masses and
SZ predictions) employ the primary models.  The models derived from
profiles with the central regions excised we identify as ``alternate''
models.  {\it Since the alternate models will more accurately measure
the gradient in the outer regions of the cluster atmosphere, we always
use these to measure the total mass in the cooling flow clusters}.  In
\S~\ref{sec:clusterprofiles} we show that failing to do this will
overestimate the total mass by 6\% on average. We will also use the
alternate baryon models to bracket the uncertainty in our SZE
predictions induced by our choice of modeling strategy.  In the tables
that follow we refer to the primary and alternate models for a given
cluster as, e.g., ``A85'' and ``A85.2''.

For all clusters without evidence for a cooling flow, we use a single
beta model fit to the profile (both for the baryon model and for
the total mass).

Generally we find that the reduced $\chi^2$ values for the fits are
not consistent with unity.  Since it is straightfoward to obtain
robust, qualitatively good fits to the azimuthal profiles given the
methods detailed above, we do not quote the $\chi^2_{\nu}$ values for
the fits.  Appendix~\ref{app:profiles} shows our fits to the cluster
radial profiles.

Our results will be summarized in \S~\ref{sec:clusterprofiles}.

\clearpage
\subsection{Spectral Analysis}
\label{sec:spectralanalysis}

In order to determine the cluster count rate in a given energy band,
it is necessary to separate the contributions of cosmic and residual
instrumental backgrounds in the image from that of the cluster.  We
investigated two methods of accomplishing this.  First, we used a
traditional approach wherein the background count rate was estimated
from an annulus encompassing the region lying between $35'$ and $40'$
from the cluster centroid (the ``ring'' method).  Second, we tried
extracting the cluster count rates directly from each band by fitting
the azimuthal profile of that band to a beta model count rate profile
(the ``$\beta$'' method) with a free constant background.  We found
the $\beta$ method to be less sensitive to variable backgrounds and
overall somewhat more stable than the ring method; indeed, for the
soft bands (R1 and R2) the ring spectra often provide meaningless
results.  Since the $\beta$ spectra also correctly interpolate the
cluster flux in masked regions and are less sensitive to the details
of the point source masking, we used this method in the analysis.

The cluster count rates were extracted from each of the R1, R2, and
R4--R7 using this strategy.  R3 contains little signal due to a Carbon
K$\alpha$ edge caused by the PSPC window and is not generally
supported by EXSAS, so we discard the data from this band.  Due to the
likelihood of the x-ray spectra being contaminated by extra cool
phases in the cluster core, we excised the inner $150 \, h^{-1} \,$
kpc of emission for clusters in which \citet{Markevitch_et_al_1998}
detect a significant central cool component of emission.  Our fiducial
cut radius is well beyond the cooling radius for most of the clusters
in our sample.  \citet{Peres_et_al_1998} obtain a cooling radius
($r_{cool} = 102^{+38}_{-27} h^{-1} \,$ kpc) for A1795, one of the
strongest cooling-flow clusters in our sample; even for this cluster,
there is a margin for error. This relatively extreme excision also
insures that our models are sampling the large scales relevant for the
SZE and baryonic mass. The fractional uncertainty in each band's count
rate was calculated by adding the Poisson uncertainty (calculated from
the equivalent non-vignetting-corrected spectrum) in quadrature with
an additional 1.5\% uncertainty estimated by inspecting the fits to a
few initial data sets.  This latter correction was necessary to make
$\chi_{\nu}^2 \sim 1$ for fits to these data.

These count rates and uncertainties were converted into {\tt .pha}
files using the {\tt FTOOL} {\tt ascii2pha} and analyzed with {\tt
XSPEC}.  Each spectrum was modeled by one or more Raymond-Smith
thermal plasmas with photoelectric absorption as described below.  The
plasma temperatures and redshifts used in this analysis are those
shown in Table~\ref{tbl:redshift_temp}.
We used the standard GSFC response
matrices\footnote{Available at {\tt
http://heasarc.gsfc.nasa.gov/docs/rosat/pspc\_matrices.html}} for the
PSPC B, PSPC B(H), and PSPC C, rebinned as per the energy channel
definitions of \citet{Snowden_et_al_1994}.

In our initial analysis of the spectral data, we allowed the absorbing
column density to be a free parameter which we then fit to the R2 and
R4--R7 data.  The results of this exercise are somewhat discouraging.
Out of 22 clusters, 4 (A478,A754, A2256, and A3266) show absorbing
column densities higher than the Galactic value, 3 (A401, A2029, and
A3158) are consistent or marginally consistent with the Galactic
value, and 15 are significantly lower than the Galactic value.  While
it would be possible for the column densities to be systematically
{\it higher} than the Galactic column density due to extragalactic
neutral hydrogen along the line of sight (e.g., associated with the
cluster), no effect can make the absorbing column densities
systematically {\it lower} than the Galactic value.  Substructure in
the Galactic disk, for example, should only cause a scatter in the
observed column densities.

The fact that we consistently observe a deficit in the fitted column
density relative to the Galactic expectation is equivalent to there
being an excess of counts in the soft channels if we consider the
Galactic hydrogen column densities, on average, to be reliable.  There
are two possible explanations for such an effect.  One is that in most
of the clusters we observe, there is a widespread cool phase.  While
this is not likely at radii larger than the cooling radius, a variety
of astrophysical mechanisms may permit such a situation \citep[see,
for example,][]{Gunn_and_Thomas_1996}.  In order to diagnose the
significance of this, we examined in detail three of the clusters
(A85, A1651, and A1656) showing particularly low column densities in
our spectral fits.  In two of the three cases (A1651 and A1656) fits
of comparable quality to the free Hydrogen column density fit are
achieved if the gas is assumed to be at $\sim 0.1$ keV with the column
density fixed at the Galactic value; in these cases, the contribution
of the cooler phase to the observed flux is $\sim 5\%$ of the total
flux.  Since the cool phases (under the assumption of hydrostatic
equilibrium) are highly over-represented in the x-rays, such a phase
in itself contributes negligibly to the baryonic mass and hence the
SZE.  On the other hand, the best-fit flux for the dominant (hot)
Raymond-Smith component is reduced by $\sim 5\%$ for these two
clusters; this is comparable to the reduction in de-absorbed flux that
we infer for a single-component fit with a free absorbing column
density for these clusters.

A more likely possibility is that there is a
systematic error in the PSPC calibration.  This has recently been
suggested by \citet{Iwasawa_et_al} on the basis of a comparison of the
spectral indices inferred from ASCA and ROSAT data on the AGN NGC5548.
These authors find that the ROSAT spectral index is significantly
steeper than the ASCA spectral index over the same energy range.  This
is consistent with the excess of soft counts which we observe.
\citet{Markevitch_and_Vikhlinin_1997} have noted that ROSAT temperatures
are consistently lower than those obtained with ASCA, GINGA, and EINSTEIN;
this is also consistent with what we see.

The most significant case of absorption {\it over} the Galactic value
in our sample is A478.  For this cluster, the level of absorption is
enough to render R1 and R2 effectively useless; nevertheless, we are
able to constrain the absorbing column density using only bands
R4--R7.  If we exclude the central $150 h^{-1} \,$ kpc of emission and
fix the absorbing column density at the Galactic value ($14.8 \times
10^{20} {\rm cm^{-2}} \,$), the best-fit Raymond-Smith model has
$\chi_{\nu}^2 = 17.7$ for 3 D.O.F.; allowing the absorbing column
density to be free and fitting for it, we obtain $21.1 \pm 1.0 \times
10^{20} {\rm cm^{-2}} \,$ with $\chi_{\nu}^2 = 1.90$ for 2 D.O.F.  We
also tried modeling the spectrum with two Raymond-Smith components
(one at 8.4 keV and one at a range of hotter temperatures) plus
Galactic absorption; all of these yielded $\chi_{\nu}^2$ values {\it
higher} than those given by the single-phase Galactic absorption fits.
Our best results are consistent with the result of a single-phase fit
to all of the cluster emission (including the cooling flow), which
yields an absorbing column density of $22.5 \pm 0.7 \times 10^{20}
{\rm cm^{-2}} \,$ and $\chi_{\nu}^2 = 3.2$ for 2 D.O.F.; allowing a
second, cool phase in this fit yields a lower $\chi_{\nu}^2$ and a
slightly higher column density.  The column densities we determine are
significantly in excess of the Galactic value.
\citet{Allen_et_al_1993} have measured an absorbing column density of
$24.9^{+1.2}_{-0.9} \times 10^{20} {\rm cm^{-2}} \,$; if we adopt the
electron temperature they have used in their analysis (6.6 keV) we
achieve results consistent with theirs at the $1.5 \sigma$ level.
Note the Allen et al. measurement also uses ROSAT data.  For this
cluster and this cluster only, we use our best-fit column density
($21.1 \pm 1.0 \times 10^{20} {\rm cm^{-2}} \,$) instead of the
Galactic column density.  If the Galactic column density is used
instead, the cluster luminosity is underestimated by 18\%.

In light of the above considerations, we have fixed all of the
hydrogen column densities (except for A478) at the Galactic value.
Most clusters we model with a single Raymond-Smith component.
However, some cluster spectra (A401, A3558, and A3921) were not well
described by a single Raymond-Smith component even when the column
density was allowed to vary.  For these clusters, including a second
plasma component at a fiducial temperature of $T_e = 1 \, {\rm keV}$
improves the fit relative to a single-phase Raymond-Smith model (even
with a free Hydrogen column density).  In all of these cases, $\sim
10\%$ of the flux is in the 1 keV component, corresponding to 1 -- 2\%
of the mass for a uniform, isobaric ICM.  It is interesting that two
of these three clusters (A401 and A3558) are in the early stages of a
major merger; the third shows a highly elliptical x-ray morphology.

For all of the clusters, the deabsorbed flux ($K$) of the hot
(dominant) Raymond-Smith component is used to determine the baryonic
mass (via Eq.~\ref{eq:density}); we use the uncertainties ($68\%$ for
one interesting parameter, $K$) determined by the {\tt XSPEC} {\tt
fit} command.  For the cases at hand these are comparable to those
determined by the {\tt error} command.  Due to the possibility of a
systematic calibration error in the soft channels, we use only R4--R7
for the spectral fits.

If we use our best fit absorbing column densities for the sample as a
whole (excluding A478), the average ratio of the de-absorbed
luminosity to the de-absorbed luminosity inferred assuming Galactic
absorption is $0.977 \pm 0.007$.  We take this $2.3\%$ error as
indicative of the level of systematic error induced by the calibration
error suggested by \citet{Iwasawa_et_al} and/or cool phases at large
radius and include it in our final error budget.

We use the fluxes $K$ determined by this method together with 
the cluster density profiles
to determine the central density $n_{eo}$ via Eq.~\ref{eq:density}.

\subsection{Notes on Individual Clusters}
\label{sec:notesonindividualclusters}

Since galaxy clusters are not entirely uniform morphologically or
spectrally, a few special cases are inevitable in the analysis of
large data sets.  Here we indicate special measures taken for
individual objects.  Unless otherwise noted, the $\chi^2$ values and
significance levels are for fits with the absorbing column density
fixed at the Galactic value and using bands R4--R7.  Note that the
clusters requiring additional soft components in the spectral fit were
identified as requiring such on the basis of fits {\it including a
free absorbing column density}, although (as indicated above, and except for A478) our
final results all assume a column density fixed at Galactic.
\begin{itemize}
\item{\bf A85:}  The emission from an infalling group of galaxies to the south of the main cluster center was excised from the image prior
to the analysis.
\item{\bf A399:}  A399 and A401 lie within the same field of view separated by some $37'$.  A401 was excised from the image (out to a radius
of $25'$) prior to the analysis.
\item {\bf A401:} An excess soft component of emission over the
best-fitting Raymond-Smith model is detected at the $3 \sigma$ level.
The best fitting single component model has $\chi_{\nu}^{2} = 7.8$,
whereas allowing an additional soft (1 keV) component reduces this to
$\chi_{\nu}^{2} = 2.7$.   We used the 2-component model and ignore the cool
component obtained in the fit since its contribution to the baryonic
mass is very small.  Similar results are obtained if R2 is included in
these fits.  A399 was excised from the image out to a radius of $25'$
prior to the analysis.
\item {\bf A478:}  A significant excess over the Galactic neutral hydrogen column density is observed.
For this cluster only, we adopt the best-fit hydrogen column density instead of the Galactic value.
\item {\bf A1656:}  The emission from an infalling group of galaxies to the south-west of the cluster was excised from the image prior to
the analysis.
\item {\bf A2029:}  A spatially variable background is seen in both PSPC exposures of this field; the field of view also
 includes (just to the north of A2029) Abell
2033.  Extended emission around and between these two 
clusters is clearly seen.  
Spectra extracted by beta model fits to the profile
are not significantly affected by this.
\item {\bf A3558:}  A soft component of emission is required in the fit .  A single-phase fit to the cluster as a whole with Galactic absorption gives $\chi_{\nu}^{2} = 11.1$ for
3 D.O.F. ; a two-component fit gives $\chi_{\nu}^{2} = 4.7$ for 2 D.O.F. and a soft component detected at the $5 \sigma$ level.  A3558 is at the center of the
Shapley supercluster, with four other Abell clusters and several smaller clusters within a $2^{\circ} \,$ radius.  
Significant emission over most of the field of view (particularly to the south-east) is clearly evident.  
\item {\bf A3571:}  Located just over $4^{\circ}.3$ ($= 8.6 h^{-1}\,$ Mpc) 
away from A3558, this cluster is at a similar redshift of $z\sim 0.04$ and appears also to be associated with the Shapley
supercluster. There is some extended emission in the field.
\item {\bf A3921:}  A soft component is required in the fit.
\end{itemize}

\subsection{Assessment of Errors}
\label{sec:assessmentoferrors}

The following three subsections describe our assessment of the
statistical and systematic uncertainties in the our models and the
model predictions (\S~\ref{sec:montecarlosimulations}) and the overall
calibration uncertainty (\S~\ref{sec:calibrationuncertainties}).

\subsubsection{Monte Carlo Simulations}
\label{sec:montecarlosimulations}

To study statistical uncertainties in the profile parameters, masses,
and SZE predictions for each cluster, the composite 0.5--2.0 keV
(R4--R7) count rate image for the longest exposure on each cluster was
smoothed with a $30''$ FWHM Gaussian.  A set of $10^3$ simulated
observations were created by multiplying the smoothed count rate image
by the averaged R4--R7 exposure maps calculated by EXSAS and adding
Poisson noise.  Each realization was then subjected to an automated
analysis designed as much as possible to mimic our actual data
reduction.  Exposure and point source masks were applied to the data,
and the centroid computed within a circular aperture with a radius
randomly varying between $2'$ and $30'$.  The center of this circle
was taken to be the emission centroid we had determined by hand
(\S~\ref{sec:profileanalysis}); a random perturbation with a
Gaussian $\sigma$ of $30''$ in both coordinates was applied to this
center to account for the uncertainty in the visually determined
luminosity peak.  The azimuthal profile was computed about this
centroid and fit to a model as described in
\S~\ref{sec:profileanalysis}.  For clusters on which we had chosen to
perform fits to the exterior data, we also evaluated these fits with
the same method.

This results in a distribution of $10^3$ profile parameterizations for
each cluster which we use to determine the 68\% confidence intervals
for $\theta_0$ and $\beta$.  Using this distribution plus the
confidence interval for the cluster luminosity obtained from the
spectral analysis, we determined confidence intervals for the central
density, baryonic mass, total mass, and baryon mass fraction.  In
calculating the total mass confidence interval, 90\% confidence
intervals for $T_e$ were translated into $1-\sigma$ error bars
assuming Gaussian statistics, and these were added in quadrature to
the error bars resulting from our Monte Carlo simulations of the
cluster profiles.  We also computed the distribution of 5.5-meter and
ACBAR beam-averaged optical depths for each cluster by generating 2-D
realizations of the profile parameterizations resulting from our Monte
Carlo analysis and convolving these with the telescope beam and
switching patterns. The uncertainty in the 5.5-meter main beam
characterization ($\sigma_{5.5-m} = 3'.12 \pm 0'.11$) was included in
these calculations; for ACBAR we assumed a $4'$ FWHM primary beam.
The 5.5-meter $\tau_{sw}$ predictions were computed with a beam throw
of $22'.12$.

We used a similar strategy to quantify the sensitivity of our results
to residual PSPC backgrounds and errors in the vignetting correction.
In each case a fiducial model with $\theta_0 = 5'.0$, $\beta = 0.720$,
and a central ``intensity'' of 50.0 counts per pixel was created as a
sky brightness template; this is characteristic of a short ($\sim 8$
ksec) exposure of a merger cluster in our sample.  For the background
study, $10\%$ of the actual background which EXSAS subtracted from the
A85a data set was then added to this image, along with a constant
background of 0.05 counts/pixel.  For the vignetting study, a
quadratic vignetting error ($\pm 5\%$ at the edges of the field of
view) was applied to the image.  In the analysis of many PSPC data
sets, \citet{Vikhlinin_et_al_1999} find no evidence for vignetting
errors of greater than$\sim 5\%$, so this is not unrealistic.  $10^4$
instantiations of this map were generated including Poisson noise;
these images were then azimuthally averaged and fit to beta models.
In all three cases the average beta model parameters were affected
by less than $0.1\%$.  This implies not only that our {\it profile}
parameters are insensitive to these potential sources of systematic
error, but also that our (beta model-derived) {\it spectra} are
unaffected by them at this level. The insensitivity of our derived
parameters to the addition of a variable background can be understood
in light of the fact that while the instrumental backgrounds are not
in general constant, neither are they correlated with a
beta model.  The insensitivity of our results to the vignetting
correction is simply due to the fact that this correction matters most
in the outer wings of the azimuthal profile, which drive the constant
term in the fit.

We also used these simulations to quantify the error in the total number of counts inferred by fitting
a beta model to the data.  We find that the fractional error in the total number of counts N is
well represented by:
\begin{equation}
\frac{\epsilon_{N}}{N} = \sqrt{ \frac{2}{N}}
\end{equation}
for the given background levels.  

\subsubsection{Calibration Uncertainties}
\label{sec:calibrationuncertainties}

In assessing the level at which systematic errors affect our results,
there are two basic issues: repeatability and the overall
(non-variable) calibration uncertainty.  Some degree of
non-repeatability from one observation to the next may be introduced
if, for instance, the PSPC gains were to change slightly and in a
systematic way across the band, which would affect our spectral fits
and hence our derived luminosities.  The overall calibration is
dependent primarily on the mirror effective areas and energy response
matrices used to generate the {\tt .rmf} and {\tt .arf} files used to
calibrate the data.

We use the following data sets to evaluate the repeatability of our
measurements: A85 A--C; A401 A,B; A754 A--D; A1795 C--E; A2142 A,C;
and A2256 A,F.  The standard deviation in the ratio of the fluxes in
the individual observations to the mean flux for each object is
$2.9\%$.  We take this to characterize the $1 \sigma$ precision of our
measurements.  There is no systematic trend discernable between PSPC
B, PSPC C, or PSPC BH exposures.

\citet{Snowden_et_al_1995} 
estimate the ROSAT PSPC absolute calibration uncertainty by comparing
the RASS fluxes to the fluxes observed in other all-sky surveys.
In the 1.5-keV band, the PSPC
results agree with the HEAO-1 results to 1\%, but with Wisconsin to
only 7\%; Snowden et al. attribute this to the Wisconsin calibration
being off by this amount.  In the $\frac{1}{4} \,$ keV band, the ROSAT
fluxes appear consistently to be $\sim 10\%$ lower than the other
two surveys'.  We take the overall absolute calibration uncertainty
to be $7\%$.

Adding our $2.9\%$ precision estimate to the 2.3\% error due to the
ambiguity in the x-ray spectrum interpretation, there is a $3.7\%$ (random) calibration
uncertainty associated with each cluster.
Adding this in quadrature with the Snowden et al. systematic calibration uncertainty
of $7\%$, we obtain an overall calibration uncertainty 
of $7.9\%$ for any given cluster.  Since the electron density
is proportional to the square root of the x-ray flux, this is also the
error in $\,H_o$ due to the x-ray calibration.

\section{Results}
\label{sec:results}

In this section we summarize the results of our analysis, beginning
with the profile parameterizations (\S~\ref{sec:clusterprofiles}).
Next we present the SZE predictions for these clusters
(\S~\ref{sec:szepredictions}).  Lastly (\S~\ref{sec:masses}), we give
baryonic masses, total masses, and mass fractions for these 22
clusters.  

\subsection{Cluster Profiles}
\label{sec:clusterprofiles}

The beta model parameters and uncertainties which result from our
analysis are summarized in Table~\ref{tbl:model_enumeration}.  This
table shows the mean of the parameter distributions for the primary
models, along with, when relevant, the parameters obtained by excising
the inner $150 \, h^{-1} \, {\rm kpc}$; these are the primary and
alternate models described in \S~\ref{sec:profileanalysis}.  The
angular radius corresponding to this cut is shown as $\theta_{cut}$,
and the number of components (beta, or beta plus Gaussian for the
cooling flow) is also shown.

It is notable that the $\beta$ values for the alternate models (those with
the central data excised) tend to be
higher than those for the primary models.  To see if this is significant, we characterize
the model slopes by 
\begin{equation}
\beta_{eff}(\theta) = \beta \, \frac{\theta^2/\theta_0^2}{1+\theta^2/\theta_0^2}.
\end{equation}
or minus $1/3$ the logarithmic derivative of the density profile at
the radius $\theta$ (see Eq.~\ref{eq:hseisobeta}).
Table~\ref{tbl:beffs} shows $\beta_{eff}$ for the 10 cooling flow
clusters evaluated at $1 \, h^{-1} {\rm \, Mpc}$ for the primary and
alternate models, as well as the ratio of these quantities $r$.  The
weighted average yields $r = 0.94 \pm 0.01$, a statistically
significant result implying that the total mass estimates depend at
the $\sim 5\%$ level on the profile modeling strategy we choose for
the cooling flow clusters.  While we will not attempt to show this
here, we find that the single beta model fits to the entire profile
for cooling flow clusters are significantly more biased than fits with
a separate component to represent the central emission. Since the fits
to the outer data presumably yield a better measure of the gas slope
in this regime, we use the alternate profiles for the total mass
calculation on these 10 clusters.

%
%
\renewcommand{\baselinestretch}{1}\small\normalsize 
\begin{table}
\begin{center}
\footnotesize
\begin{tabular}{l l l l l } \\ \hline\hline
Model   & $\theta_{cut} \,$ & $N_{comp}$ & $\theta_0$ & $\beta$   \\ 
        &  $(')$              &            &   $(')$  &  \\ \hline
A85     & 0 & 2 & $2.04 \pm 0.52$ & $0.60\pm0.05$   \\
A85.2   & 3.62& 1 & $5.42\pm0.38$ & $0.779\pm0.025$   \\ \hline
A399    & 0   & 1 & $4.33  \pm 0.45 $ & $0.742 \pm 0.042$   \\\hline
A401    & 0  & 1 & $2.26 \pm 0.41 $ & $0.636 \pm 0.047$  \\\hline
A478    & 0 & 2 & $ 1.00 \pm 0.15 $ & $0.638 \pm 0.014$   \\
A478.2  & 2.22& 1 & $1.58 \pm 0.20$ & $0.683 \pm 0.011$    \\\hline  
A754    &  0  & 1 & $5.50 \pm 1.10 $ & $ 0.713\pm 0.120$   \\\hline
A780    &  0 & 2& $1.64  \pm 0.38 $ & $ 0.629\pm 0.028$    \\
A780.2  & 3.60 & 1 & $ 0.90 \pm 0.33$ & $0.640 \pm 0.007$ \\\hline
A1651   &  0 & 2 & $2.16  \pm  0.36$ & $0.712 \pm 0.036$ \\
A1651.2 & 2.39& 1 & $1.86 \pm 0.34$ & $0.690 \pm 0.020$   \\\hline
A1656   & 0   & 1 & $9.32  \pm 0.10 $ & $0.670 \pm 0.003$ \\\hline
A1795   & 0 & 2 & $2.17  \pm  0.28$ & $0.698 \pm 0.017$ \\
A1795.2 & 3.10& 1 & $2.98 \pm 0.20$ & $0.750 \pm 0.011$  \\\hline
A2029   & 0 & 2 & $0.93  \pm  0.09 $ & $0.601 \pm 0.030$\\
A2029.2 & 2.55& 1 & $ 2.09 \pm 0.36$ & $0.667 \pm 0.016$  \\\hline
A2142   & 0 & 2 & $1.60  \pm 0.12 $ & $0.635 \pm 0.012$\\
A2142.2 & 2.22& 1 & $1.91 \pm 0.60$ & $0.655 \pm 0.030$  \\\hline
A2244   & 0   & 1 &$0.82  \pm  0.14$ & $0.580 \pm 0.018$  \\\hline
A2255   & 0   & 1 & $4.36  \pm 0.12 $ & $0.723 \pm 0.015$  \\\hline
A2256   & 0   & 1 &$5.49  \pm 0.21 $ & $0.847 \pm 0.024$   \\\hline
A2597   & 0 & 2 & $0.49  \pm 0.03$ & $0.626 \pm 0.018$\\
A2597.2 & 2.33& 1 & $1.61 \pm 0.52$ & $0.693 \pm 0.022$  \\\hline
A3112   & 0 & 1 & $0.52  \pm 0.05 $ & $0.560 \pm 0.008 $\\
A3112.2 & 2.75& 1 & $ 1.24\pm 0.49$ & $0.590 \pm 0.017$  \\\hline
A3158   & 0   & 1 & $2.84  \pm 0.16  $ & $ 0.649\pm 0.018$  \\\hline
A3266   & 0   & 1 & $8.50  \pm 0.27$ & $0.942 \pm 0.020$  \\\hline
A3558   &  0  & 1 & $2.66  \pm  0.07$ & $0.55 \pm 0.006$  \\\hline
A3571   & 0 & 2 & $3.64  \pm  0.18$ & $0.669 \pm 0.009$\\
A3571.2 & 4.6 & 1 & $4.35 \pm 0.50$ & $0.702 \pm 0.020$  \\\hline
A3667   & 0   & 1 & $4.29  \pm  0.96$ & $0.589 \pm 0.051$  \\\hline
A3921   &  0  & 1 & $1.33  \pm  0.23$ & $ 0.541 \pm 0.031$  \\\hline
\end{tabular}
\end{center}
\caption{Enumeration of the primary and (for cooling flow clusters)
secondary models employed in the analysis. Errors are 68\% confidence
intervals.}
\label{tbl:model_enumeration}
\end{table}
\renewcommand{\baselinestretch}{1.6}\small\normalsize \phantom{z}

%
%
\renewcommand{\baselinestretch}{1}\small\normalsize 
\begin{table}[h]
\begin{center}
\begin{tabular}{ l l l l}
\hline\hline
Cluster & \multicolumn{2}{c}{$\beta_{eff}(1 \, h^{-1} \, {\rm Mpc}) \,$} & $r$ \\
         & All  & Outer & \\ \hline
A85      &  $0.60 \pm 0.05$ & $ 0.74 \pm 0.02 $ & $0.81 \pm 0.07$\\
A478     & $0.64 \pm 0.01$   & $ 0.67 \pm 0.01 $& $0.96 \pm 0.02$ \\
A780     & $ 0.63 \pm 0.03 $ & $ 0.63 \pm 0.01$& $1.00 \pm 0.05$ \\
A1651    & $0.70 \pm 0.03$ & $0.68 \pm 0.02$& $1.03 \pm 0.05$ \\
A1795    & $ 0.69 \pm 0.01 $ & $0.73 \pm 0.01$& $0.94 \pm 0.02$ \\
A2029    & $ 0.60 \pm 0.01 $ & $ 0.66 \pm 0.02 $& $ 0.91 \pm 0.03$\\
A2142    & $ 0.63 \pm 0.01 $ & $ 0.65 \pm 0.02$& $ 0.97 \pm 0.03$\\
A2597    & $0.62 \pm 0.02$ & $0.68 \pm 0.02$& $0.91 \pm 0.04$ \\
A3112    & $0.56 \pm 0.01 $ & $0.60 \pm 0.01$& $0.93 \pm 0.02 $\\
A3571    & $0.66 \pm 0.01$  & $0.69 \pm 0.02$& $0.96 \pm 0.03$ \\\hline
AVERAGE  &                  &                & $0.94 \pm 0.01$ \\\hline
\end{tabular}
\end{center}
\caption{$\beta_{eff} \,$ as a function of model choice for the 10 cooling flow clusters
in our sample. The ``All'' models are beta plus Gaussian fits to the entire profile; the ``Outer''
models are beta model fits to the outer data only.  Errors are 68\% confidence
intervals.}
\label{tbl:beffs}
\end{table}
\renewcommand{\baselinestretch}{1.6}\small\normalsize \phantom{z}

\clearpage

\subsection{Sunyaev-Zeldovich Effect Models and Predictions}
\label{sec:szepredictions}

The inverse Compton optical depths we compute for the 5.5-meter and
ACBAR are displayed in Table~\ref{tbl:tau_conf}; for clusters with a
central excess we show predictions using both the primary and
secondary models. The fitting problem is somewhat complicated in that
there are many characteristic scales; some of them are angular, such
as the beam width and chop, and some metric, such as the excision
radius and any intrinsic cluster scales.  The accuracy with which the
beam-averaged and switched optical depths are predicted depends on
this hierarchy of scales as well as the intrinsic signal-to-noise
level of the data.  Nevertheless some overall trends are apparent.

\begin{figure}
\vspace{3.8in}
\includegraphics{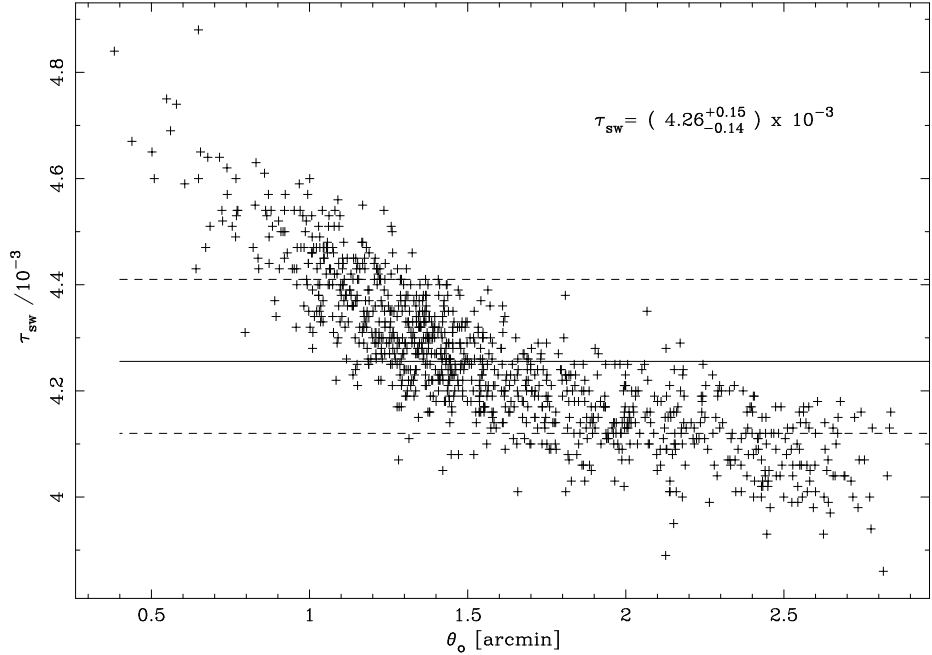}
\caption{Switched 5.5-m SZE as a function of core radius for fits to A2142 excluding the inner $150 \, h^{-1} \,$ kpc
showing that so long as parameter degeneracies are correctly accounted for and a reasonable fit obtained, our results are
insensitive to the form the profile at the $\sim 3\%$ level. The mean and 68\% confidence increments are 
shown as solid and dashed lines, respectively; the standard deviation in this distribution (which excludes the 
uncertainty in $\Omega_{Beam} \,$ for the 5.5-m) is 2.3\% of the mean.}
\label{fig3}
\end{figure}

First, the inverse Compton optical depths are predicted to
significantly better accuracy than might be naively expected on the
basis of the profile analysis, the parameters of which are often
uncertain at the $ 10\%$ level or greater (see
Table~\ref{tbl:model_enumeration}).  In contrast to this the 5.5-meter
optical depths in Table~\ref{tbl:tau_conf} are predicted with an
average accuracy (for the 5.5-meter $\tau_{sw}$) of 5.4\%; if the main
beam uncertainty is excluded from our simulations, the average
uncertainty is $\sim 2\%$.  The fact that the models can be
constrained more accurately than the individual model parameters is
caused by the well-known parameter degeneracies inherent in the beta
model analysis.  To further illustrate this point, we show in
Figure~\ref{fig3} the distribution of $\tau_{sw}$ for the 5.5-meter
for the alternate model A2142.2 (excluding, however, the main beam
uncertainty).  Due to the central excision, the core radius in this
case is very poorly constrained; even so, $\tau_{sw}$ is predicted to
an accuracy of $2.3\%$. It is notable that the uncertainties on the
ACBAR optical depths are typically a factor of $\sim 2$ smaller than
those on the 5.5-meter beam-averaged optical depths.  The reason for
this is that we have not included in the ACBAR beam uncertainty in our
simulations, since this is not yet known.  The discrepancy is most
notable in the cases where the clusters subtend a large angle on the
sky, such as A754 and A2256; for these two cases, the uncertainty in
$\tau_{Beam}$ for the 5.5-meter is more than three times that in
$\tau_{Beam}$ for ACBAR.  The 5.5-meter primary beam uncertainty is in
part responsible for the larger statistical uncertainties for the A85,
A754, and A780 models, as these clusters all lie at $z < 0.06$.  The
beam uncertainty dominates the statistical error in $\tau_{Beam}$ and
$\tau_{sw}$ for the Coma cluster (A1656), which to a good
approximation fills the main beam of the 5.5-meter telescope.

Second, the uncertainties in $\tau_{sw}$ are often {\it less} than
those in the associated $\tau_{Beam}$.  This occurs because the noise
in $\beta$ is suppressed somewhat in the switched SZE predictions.
For a fixed x-ray flux, increasing $\beta$ reduces the predicted
central decrement by reducing the baryonic mass surface density.  At
the same time, this increase reduces the flux in the chopping beam,
thereby increasing the switched decrement and cancelling some of the
reduction due to the lesser central decrement.  A similar effect will
be present for interferometers, which effectively measure sky
temperature differences on the angular scale of the fringe pattern.

Partly due to this effect, and partly due to the fact that the
5.5-meter beam is larger than the excision radius for most of these
clusters, our $\tau_{sw}$ results for the 5.5-meter are statistically
unaffected by our chosen modeling strategy.  All of the $\tau_{sw}$
predictions for the alternate models in Table~\ref{tbl:tau_conf} are
consistent (or, for A780, marginally consistent) with the primary
model predictions.  This is certainly not the case for $\tau_o$, for
which the two strategies typically yield $\tau_o$ estimates which
differ by $\sim 30\%$ or more.  For clusters at $z < 0.051$, the
excised region of the profile is equal to or larger than the diameter
of the 5.5-meter primary beam.  This explains the relatively large
discrepancy between A780 and A780.2 (a $z=0.0522$ cluster).  Note that
even in these cases, the alternate model will, under the assumption of
spherical symmetry, correctly measure the contributions from the
extended lines of sight in the beam.

The ACBAR beam-averaged optical depths are somewhat less robust in
this respect.  This is not surprising because the $150 \, h^{-1} \,
{\rm kpc}$ excision radius is larger than the ACBAR primary beam for
all of the clusters in this sample.  Better results could certainly be
obtained with a more moderate excision.  Even though we have not
explicitly demonstrated that these results are robust as we have for
the 5.5-meter predictions, the primary models we present here should
be more than adequate for experiments like ACBAR with $\sim 4'$
angular resolution given the beam uncertainties and measurement errors
likely to exist in practice.

Abell 754 is the cluster with the least accurate profile.  This is a
well studied major merger cluster
\citep[e.g.,][]{1986ApJ...308..530F,1995ApJ...443L...9H,1996ApJ...466L..79H}
with clear large-scale substructure in both x-ray and optical images.
We include it in the current analysis for completeness, although any
total mass estimates for this cluster should be interpreted with
caution.  Even for this cluster, the $10\%$ uncertainty in $\tau_{sw}$
is only just comparable to typical SZ measurement errors.

%
%
\renewcommand{\baselinestretch}{1}\small\normalsize 
\begin{table}[h!]
\footnotesize
\begin{center}
\begin{tabular}{l l l l l l}
\hline\hline
Cluster       &       & \multicolumn{2}{l}{\phantom{tbeamyy}5.5-meter}   & ACBAR \\
Model         & $\tau_o$      & $\tau_{beam} \,$ & $\tau_{sw} \,$ & $\tau_{beam} \,$ \\ 
                & $(10^{-3} \, h^{-1/2})$ & $(10^{-3} \, h^{-1/2})$ & $(10^{-3} \, h^{-1/2})$ & $(10^{-3} \, h^{-1/2})$ \\ \hline
A85          & $  6.60 \pm 0.48   $ &  $ 3.81 \pm 0.22  $    &  $ 2.88 \pm 0.23 $      &  $  4.94 \pm 0.12$ \\
A85.2      & $4.72 \pm 0.14$ & $3.45 \pm 0.09   $          & $  2.76 \pm 0.13    $     & $4.14 \pm 0.10$  \\ \hline
A399         & $ 4.34 \pm 0.28    $ &  $ { 3.08 \pm 0.17}  $    &  $ 2.50 \pm 0.13  $      &  $  3.77 \pm 0.05 $  \\\hline
A401         & $6.96 \pm 0.55  $ &  $ 4.15 \pm 0.24 $     &  $ 3.17 \pm 0.18  $    &  $5.34 \pm 0.13 $    \\\hline
A478         & $ 12.93 \pm 1.17    $ &  ${ 4.39 \pm 0.20}  $     &  $3.68 \pm 0.15  $    &   $ 6.59 \pm 0.12  $  \\
A478.2     &  $10.02 \pm 1.03$   & $  {4.22 \pm 0.22}    $           &    $   3.61 \pm 0.17   $     &  $6.15 \pm 0.23$         \\\hline
A754         & $ 5.20 \pm 0.37  $ &  $  4.16 \pm 0.27$     &  $ 3.02 \pm 0.31  $    &  $ 4.77 \pm 0.09  $  \\\hline
A780         & $5.49^{+1.18}_{-0.54} \,$ &  $ 2.62 \pm 0.13 $      &  $ 2.11 \pm 0.11 $    &    $ 3.61 \pm 0.05 $  \\
A780.2       & $10.67^{+2.42}_{-2.97} \,$ &  $ 2.98 \pm 0.18 $      &  $ 2.52 \pm 0.20 $    &    $ 4.57 \pm 0.45 $  \\\hline
A1651        & $5.75^{+0.80}_{-0.51} \,$ &  $2.84 \pm 0.15   $      &  $ 2.44 \pm 0.11 $    &    $3.98^{+0.20}_{-0.09} \,$  \\
A1651.2    & $6.06 \pm 0.84 $ &  $ 2.88 \pm 0.12  $      & $  2.47 \pm 0.11 $   &    $ 4.07 \pm 0.23   $  \\\hline
A1656        & $5.22 \pm 0.36$ &  ${ 4.77 \pm 0.30}  $     &  $ 2.76 \pm 0.16  $    &    $ 5.07 \pm 0.02  $  \\\hline
A1795        & $6.88 \pm 0.54$ &  $ 3.49 \pm 0.17  $     &  $ 2.95 \pm 0.13  $    &   $ 4.84^{+0.08}_{-0.12}  $  \\
A1795.2    & $ 5.94 \pm 0.29$ &  $ 3.37 \pm 0.11  $      &  $ 2.88 \pm 0.10  $    &   $4.52 \pm 0.13$  \\\hline
A2029        & $ 12.13 \pm 1.31    $ &  $4.37 \pm 0.20  $      &  $3.52 \pm 0.14   $    &   $ 6.34 \pm 0.10 $  \\
A2029.2         & $8.09 \pm 1.51 $ &  $4.09 \pm 0.19  $      &  $3.37 \pm 0.17  $    &  $5.60 \pm 0.43$  \\\hline
A2142        & $ 11.11 \pm 0.09$ &  $ {5.28 \pm 0.26}  $     &  $ 4.28 \pm 0.18  $    &   $ 7.33 \pm 0.06  $  \\
A2142.2    & $10.68 \pm 2.13$ &  ${5.19 \pm 0.30}  $      &  $ 4.25 \pm 0.22 $    &   $7.16 \pm 0.50$  \\\hline
A2244        & $ 8.16 \pm 0.64$ &  $2.93 \pm 0.15 $    &  $2.30 \pm 0.11$      &  $ 4.20 \pm 0.07 $  \\\hline
A2255        & $ 3.89 \pm 0.27$ &  $2.81 \pm 0.15 $    &  $ 2.23 \pm 0.11   $      &  $3.40 \pm 0.03   $  \\\hline
A2256        & $5.13 \pm 0.36$ &  $ {3.77 \pm 0.20}  $    &  $3.19 \pm 0.16   $      &  $  4.54 \pm 0.03 $  \\\hline
A2597        & $9.86 \pm 0.75$ &  $2.05 \pm 0.10$    &  $ 1.74 \pm 0.09 $      &  $3.26 \pm 0.09$  \\\hline
A3112        & $9.04 \pm 0.65 $ &  $2.62 \pm 0.13$    &  $2.06 \pm 0.10$      &  $3.82 \pm 0.06  $  \\
A3112.2         & $6.32^{+2.11}_{-1.34} $ &  $ 2.58 \pm 0.13  $      &  $2.05 \pm 0.14  $    &  $3.63 \pm 0.30$ \\\hline
A3158        & $4.88 \pm 0.34$ &  $3.11 \pm 0.16$    & $2.43 \pm 0.12$      &  $3.95 \pm 0.43 $  \\\hline
A3266        & $4.30 \pm 0.30$ &  $ 3.56 \pm 0.20$    &  $ 2.89 \pm 0.16$      &  $4.03 \pm 0.03$  \\\hline
A3558        & $5.07 \pm 0.36$ &  $3.59 \pm 0.20   $ &  $ 2.39 \pm 0.13 $      &  $4.31 \pm 0.08$  \\\hline
A3571        & $6.46 \pm 0.47$ &  $4.52 \pm 0.25$    &  $3.49 \pm 0.17$      &  $5.54 \pm 0.10$  \\
A3571.2         & $6.01 \pm 0.29 $ &  $  4.39 \pm 0.16 $      &  $  3.41 \pm 0.14$    &   $5.27 \pm 0.02$  \\\hline
A3667        & $5.60 \pm 0.46$ &  $ 4.52 \pm 0.28  $    &  $ 2.95 \pm 0.17$      &  $5.18 \pm 0.17$  \\\hline
A3921        &  $5.75 \pm 0.41 $     &  $ 3.06 \pm 0.16 $    &  $2.18 \pm 0.12$      &  $ 3.98 \pm 0.04$  \\\hline
\end{tabular}
\end{center}
\caption{Central and beam-averaged optical depths for 22 clusters.  The 5.5-meter beam-averaged
optical depths include the uncertainties due to the main beam determination.  All quoted 
$\tau$ values are the means of the observed distributions, {\it not} the values corresponding to our
best-fit models.  The models are defined in Table~\ref{tbl:model_enumeration}.
All errors are $1-\sigma$.}
\label{tbl:tau_conf}
\end{table}
\renewcommand{\baselinestretch}{1.6}\small\normalsize \phantom{z}

Our best fit profile models are summarized in
Table~\ref{tbl:best_models}.  Due to the beta model parameter
degeneracies, it is important to use the parameters listed in
Table~\ref{tbl:best_models} and not, for instance, the distribution
mean for $n_{eo}$ from Table~\ref{tbl:mass_distr} together with the
$\beta\,$ and $\theta_0$ from Table~\ref{tbl:model_enumeration}.  The
values for $\tau_0$ shown in Table~\ref{tbl:best_models} are model
normalizations and are {\it not} necessarily representative of the
actual central inverse Compton optical depth which would be observed
in very high resolution SZ maps.

The dependence of these results on the chosen cosmology will be discussed
in \S~\ref{sec:conclusion}.

%
%
\renewcommand{\baselinestretch}{1}\small\normalsize 
\begin{table}
\begin{center}
\begin{tabular}{ l l l l l l }
\hline\hline
Cluster    &$\theta_0$  & $R_{0}   $   & $\beta$ & $n_{eo} \,$ & $\tau_o$ \\
           & $(')$  & $ (h^{-1} \, {\rm kpc})$   &  &($10^{-3} \, h^{1/2} \, {\rm cm^{-3}} $)  & ($10^{-3} \, h^{-1/2}$)  \\\hline
A85        &  $ 2.04  $ & 84.4     &  $0.600   $    &  $ 10.44  $   &  $ 6.59  $  \\
A399       &   $ 4.33 $ &  239     &  $ 0.742  $    &  $ 3.23  $   &  $ 4.34  $ \\
A401       &  $  2.26 $ & 130.     &  $0.636   $    &  $ 7.90  $   &  $  7.05 $  \\
A478       &  $ 1.00  $ & 67.5     &  $ 0.638  $    &  $ 27.81  $   &  $ 12.84  $  \\
A754       &  $ 5.50  $ & 232      &  $  0.713 $    &  $ 3.81  $   &  $ 5.20  $  \\
A780       &  $ 1.64  $ & 68.3     &  $ 0.629  $    &  $ 11.30  $   &  $ 5.38  $  \\
A1651      &  $ 2.16  $ & 135      &  $ 0.712  $    &  $ 7.14  $   &  $ 5.70  $  \\
A1656      &  $ 9.32  $ & 181      &  $ 0.670  $    &  $4.52   $   &  $ 5.23  $  \\
A1795      &  $2.17   $ & 105      &  $ 0.698  $    &  $ 10.70  $   &  $ 6.81  $  \\
A2029      &  $  0.930$ & 54.7     &  $ 0.601  $    &  $  29.35 $   &  $ 11.99  $  \\
A2142      &  $ 1.60  $ & 108      &  $0.635   $    &  $ 14.95  $   &  $ 11.10  $  \\
A2244      &  $ 0.820 $ & 59.5     &  $ 0.580  $    &  $ 17.30  $   &  $ 8.13  $  \\
A2255      &  $ 4.36  $ & 266      &  $ 0.723  $    &  $2.53   $   &  $ 3.89 $  \\
A2256      &  $  5.49 $ & 260      &  $ 0.847  $    &  $ 4.08  $   &  $ 5.13  $  \\
A2597      &  $ 0.49  $ & 31.7     &  $  0.626 $    &  $ 44.64  $   &  $ 9.90  $  \\
A3112      &  $  0.52 $ & 28.3     &  $ 0.56  $    &  $ 38.12  $   &  $ 9.02  $  \\
A3158      &  $  2.84 $ & 132      &  $ 0.649  $    &  $ 5.52  $   &  $ 4.88  $  \\
A3266      &  $ 8.50  $ & 398      &  $0.942   $    &  $ 2.49  $   &  $ 4.30  $  \\
A3558      &  $ 2.66  $ & 103      &  $ 0.55  $    &  $ 5.71  $   &  $ 5.07  $  \\
A3571      &  $ 3.64  $ & 118      &  $0.669   $    &  $ 8.57  $   &  $ 6.46  $  \\
A3667      &  $  4.29 $ & 188      &  $ 0.589  $    &  $ 3.92  $   &  $  5.68 $  \\
A3921      &  $  1.33 $ & 94.9     &  $ 0.541  $    &  $ 6.18  $   &  $  5.20 $  \\
\hline
\end{tabular}
\end{center}
\caption{Summary of best-fit models and the central densities and optical depths
for these models. }
\label{tbl:best_models}
\end{table}
\renewcommand{\baselinestretch}{1.6}\small\normalsize \phantom{z}

\clearpage
\subsection{Baryonic and Total Masses and Baryon Mass Fractions}
\label{sec:masses}

The baryonic masses, total masses, and baryon fractions for the 22 clusters with public data in our sample are shown in 
Table~\ref{tbl:mass_distr} and Table~\ref{tbl:mass_distr_2}.  
The baryonic mass within a spherical radius $R$ for a beta model can be shown
to be:
\begin{equation}
M_{bary} (< R)= \frac{4}{3} \pi R^3 \rho_B \, \,_{2} \! F_{1}\left(\frac{3}{2},
\frac{3 \beta}{2},
\frac{5}{2},-(\frac{R}{r_o})^2\right)
\end{equation}
where $\rho_B$ is the baryonic mass density  
\begin{equation}
\rho_B = n_e \, m_p \times \mu_e .
\end{equation}
Here $m_p$ is the proton mass and $\mu_e$ is the baryonic mass in the
plasma per electron, and $ \,_{2} \! F_{1}$ is a confluent
hypergeometric function.  The total masses were computed under the
assumption of hydrostatic equilibrium.  For the total mass contained
within a sphere of radius $r$ we find
\begin{equation}
\label{eq:hse}
M(<r) = - \frac{k T(r)}{G \mu m_p} \times r \times \left( \frac{\partial \ln n_e}{\partial
 \ln r} + \frac{\partial \ln T}{\partial \ln r} \right),
\end{equation}
where $\mu$ is the mean molecular mass ($=0.592$ for our assumed 30\%
metallicity and the Solar abundances of
\citet{Anders_and_Grevesse_1989}).  For the isothermal beta model this
assumes the form
\begin{equation}
\label{eq:hseisobeta}
M(<r) = \, \frac{k T}{G \mu m_p} \times 3\, \beta \, \frac{\theta^2/\theta_0^2}{1+\theta^2/\theta_0^2}.
\end{equation}

We compute these quantites both at a fixed metric radius of $500 \,
h^{-1} \, {\rm kpc}$ and at the radius $R_{500}$ within which the mean
density is 500 times the critical density.  For the latter, we use the
\citet{Mohr_et_al_1999} definition
\begin{equation}
R_{500} = 1.185 \, h^{-1} \, {\rm Mpc} \, \times \left( \frac{T_e}{10 \, {\rm keV}} \right)^{\frac{1}{2}}.
\end{equation}
This allows us to compare virially similar regions of each cluster, as
well facilitating a direct comparison between our results and those of
Mohr et al. (although the electron temperatures they adopt tend to be
lower than ours).

Within $500 \, h^{-1} \, {\rm kpc}$, the baryonic masses are accurate
to $4\%$ or better, with an average uncertainty of about 2\%.  The
mean baryonic mass within this radius is $(1.42 \pm 0.18)\, \times
10^{13}\, h^{-5/2} \, \msun$ ($1\sigma$ S.D.). The total masses are
accurate to $5 - 10\%$, with an average uncertainty of $8\%$; the mean
total mass inside $500 \, h^{-1} \, {\rm kpc}$ is $(2.19 \pm 0.36) \,
\times 10^{14} \, h^{-1} \, \msun$.  At $R_{500}$, the average
baryonic mass is $(2.94 \pm 0.85)\, \times 10^{13}\, h^{-5/2} \,
\msun$, and the average total mass is $(4.72 \pm 1.86) \, \times
10^{14} \, h^{-1} \, \msun$.  The mean baryon fraction within $500 \,
h^{-1} \, {\rm kpc} \, $ is $(6.09 \pm 0.17) \, h^{-3/2}\,\% \, ({\rm
s.d.} = 0.78 \,h^{-3/2} \,\%) \,$; within $R_{500}$, it is $ (7.02 \pm
0.28 ) \, h^{-3/2}\,\% \, ( {\rm s.d. } = 1.32\, h^{-3/2}\, \%)$.

As noted in \S~\ref{sec:profileanalysis}, the total masses we infer
depend at the 5\% level on the profile modeling strategy we choose,
particularly for the cooling flow clusters.  For these clusters we
compute the total mass from fits to the data excluding the inner
region of the cluster.  The baryonic masses (like the SZE predictions
in in \S~\ref{sec:szepredictions}) are always computed from single- or
double-component fits to all of the profile.

The total mass computation relies on the assumptions of hydrostatic
equilibrium and isothermality.  \citet{Evrard_et_al_1996} find that
isothermal beta model estimates of the total mass in numerically
simulated clusters on scales of $R_{500} - R_{2000}$ are unbiased and
accurate to $\sim 15\%$ on average, suggesting that the strategy we
adopt is not likely to be seriously in error.
\citet{Markevitch_et_al_1998} have found evidence for departures from
isothermality in their analysis of 30 nearby clusters.  In a similar
analysis, however, \citet{Irwin_et_al_1999} do not find a similar
effect.  If the temperature were to fall at large radius, the total
masses would be lower than the estimates we present here.  Future data
from XMM and Chandra will help clarify this situation.

We will discuss the effects of the chosen cosmological model on these
results in the next secion.

%
%
\renewcommand{\baselinestretch}{1}\small\normalsize 
\begin{table}[h!]
\footnotesize
\begin{center}
\begin{tabular}{l l l l l }
\hline\hline
Cluster  &  $M_{Bary} \,$  & $M_{tot} \,$   & $f$ &  $n_{eo} \,$   \\ 
         & $(10^{13} \, h^{-5/2} \, \msun)$ & $(10^{14} \, h^{-1} \, \msun)$ &$( h^{-3/2} \, \%)$ & $(10^{-3} \, h^{1/2} \, {\rm cm^{-3}})$ \\ \hline
A85   (500 $\, h^{-1} \,$ kpc)     & $1.27 \pm 0.07  $       &   $2.26^{+0.18}_{-0.23} \,$     &   $5.61 \pm 0.21$   &  $11.14^{+2.54}_{-1.98} \,$ \\
($R_{500} =0.99 h^{-1} \,$ Mpc )&     $3.20 \pm 0.07$ &   $5.66 \pm 0.27$  &  $5.67\pm0.24$  & ---    \\

A399 (500 $\, h^{-1} \,$ kpc)       & $ 1.33 \pm 0.02 $     &  $ 2.34 \pm 0.10 $     &  $ 5.71 \pm 0.20 $       & $ 3.24^{+0.14}_{ - 0.19}  $     \\ 
($R_{500}=0.99 h^{-1} \,$ Mpc)              & $  3.51 \pm 0.08 $       &  $  5.36 \pm 0.32 $       &  $  6.55 \pm 0.32  $      & --- \\

A401 (500 $\, h^{-1} \,$ kpc)       & $1.67\pm0.05$     &  $2.64^{+0.18}_{-0.11} \,$     &  $6.35\pm 0.23$      & $8.01^{+0.56}_{-1.02} \,$       \\ 
($R_{500} = 1.06h^{-1} \,$ Mpc )              &   $4.64 \pm 0.14$     &  $5.86^{+0.50}_{-0.32} \,$      &  $7.94^{+0.44}_{-0.62} \,$      & --- \\

A478 (500 $\, h^{-1} \,$ kpc)       &  $2.01 \pm 0.03$    & $2.92^{+0.18}_{-0.33} \,$      &  $6.91^{+0.69}_{-0.40} \,$      &  $28.9^{+15.2}_{-3.9} \,$      \\ 
($R_{500} = 1.08 h^{-1} \,$ Mpc)              &  $5.17 \pm 0.25$      &  $6.81^{+0.49}_{-0.78} \,$      &   $7.60^{+0.76}_{-0.45} \,$     & --- \\

A754 (500 $\, h^{-1} \,$ kpc)       &  $1.56^{+0.10}_{-0.17} \,$    & $3.09^{+0.36}_{-0.48} \,$      &  $5.07 \pm 0.26$      &  $3.73^{+0.07}_{-0.1} \,$      \\ 
($R_{500} = 1.15 h^{-1} \,$ Mpc)              &    $5.15 \pm 0.24$    & $8.30^{+1.41}_{-1.82} \,$    & $6.21^{+1.81}_{-1.23} \,$       & --- \\

A780 (500 $\, h^{-1} \,$ kpc)       &  $0.870^{+0.014}_{-0.053} \,$    & $1.47^{+0.09}_{-0.19} \,$      &  $5.91 \pm 0.36 $      & $   12.90^{+18.0}_{-2.97} \,$    \\ 
($R_{500} = 0.77 h^{-1} \,$ Mpc)               &    $ 1.50\pm 0.05$   & $2.35 \pm 0.15$       &  $6.41 \pm 0.36$      & --- \\

A1651 (500 $\, h^{-1} \,$ kpc)      &   $1.31 \pm 0.02$   & $2.09^{+0.10}_{-0.19} \,$      &  $6.27 \pm 0.28$      &  $7.47^{+5.17}_{-0.96} \,$      \\ 
($R_{500} = 0.92 h^{-1} \,$ Mpc)              &    $2.79 \pm 0.07$    &  $4.29 \pm 0.22$      &  $6.53 \pm 0.30$      & --- \\

A1656 (500 $\, h^{-1} \,$ kpc)      &  $1.44 \pm 0.01$    & $2.99 \pm 0.14  $ & $  4.82 \pm 0.21 $            &  $4.51 \pm 0.04$ \\
($R_{500} = 1.13 h^{-1} \,$ Mpc)              &  $4.48 \pm 0.12$      &  $ 7.41 \pm 0.41 $      &   $6.03 \pm 0.28  $     & --- \\

A1795 (500 $\, h^{-1} \,$ kpc)      &   $1.34 \pm 0.01$   & $2.89 \pm 0.24$      &  $4.66\pm 0.35$  &  $11.29^{+0.61}_{-1.77} \,$      \\ 
($R_{500}= 1.04 h^{-1} \,$ Mpc)              &   $3.16 \pm 0.14$     &  $6.63 \pm 0.61$      &   $4.77 \pm 0.36$     & --- \\

A2029 (500 $\, h^{-1} \,$ kpc)      &  $1.77 \pm 0.02   $    &  $ 3.00 \pm 0.22  $     &  $ 5.90 \pm 0.38   $      &    $ 31.11^{+7.94}_{-6.40} \,$    \\ 
($R_{500} = 1.13 h^{-1} \,$ Mpc)              &    $5.01 \pm 0.21  $    &   $ 7.48 \pm 0.62 $     &  $6.70 \pm 0.46  $      & --- \\

A2142 (500 $\, h^{-1} \,$ kpc)      &   $2.38 \pm 0.02$   & $3.25^{+0.32}_{-0.24} \,$      &  $7.30^{+0.49}_{-0.67} \,$      & $15.03^{+0.92}_{-1.07} \,$       \\ 
($R_{500}= 1.16 h^{-1} \,$ Mpc )              &  $7.19 \pm 0.37$     &  $8.08^{+0.95}_{-0.73} $      &  $8.90^{+0.64}_{-0.91} \,$      & --- \\
\hline
\end{tabular}
\end{center}
\caption{Central densities, total masses, and baryon fractions for 11 clusters at
$500 h^{-1} \,$ kpc and $R_{500} \,$. Quoted densities are the mean for the distribution, {\it not} the value corresponding to our quoted
best-fit model.  Errors are $1-\sigma$.}
\label{tbl:mass_distr}
\end{table}
\renewcommand{\baselinestretch}{1.6}\small\normalsize \phantom{z}

\clearpage

%
%
\renewcommand{\baselinestretch}{1}\small\normalsize 

\begin{table}[h!]
\footnotesize
\begin{center}
\begin{tabular}{l l l l l }
\hline\hline
Cluster  &  $M_{Bary} \,$  & $M_{tot} \,$   & $f$ &  $n_{eo} \,$   \\ 
         & $(10^{13} \, h^{-5/2} \, \msun)$ & $(10^{14} \, h^{-1} \, \msun)$ &$( h^{-3/2} \, \%)$ & $(10^{-3} \, h^{1/2} \, {\rm cm^{-3}})$ \\ \hline
A2244 (500 $\, h^{-1} \,$ kpc)      & $1.31  \pm 0.03   $     &  $  2.25^{+1.07}_{-0.52} $     &  $5.84^{+1.32}_{-2.75}  $       & $ 17.73^{+1.95}_{-2.65}   $       \\ 
($R_{500} = 0.99 h^{-1} \,$ Mpc)              & $  3.33 \pm 0.37  $       &  $ 4.51^{+2.19}_{-1.12}  $       &  $  7.39^{+1.70}_{-3.49} $      & --- \\
A2255 (500 $\, h^{-1} \,$ kpc)      & $1.23 + 0.01  $     &  $ 2.27^{+0.62}_{-0.30} $     &  $ 5.41^{+0.86}_{-1.65} $       & $ 2.52 \pm 0.03  $       \\ 
($R_{500} = 1.01 h^{1} \,$ Mpc)              & $ 3.54 \pm 0.26  $       &  $  5.51^{+1.51}_{-0.73} $       &  $ 6.43^{+0.85}_{-1.76}  $      & --- \\
A2256 (500 $\, h^{-1} \,$ kpc)      & $1.62 \pm  0.01   $     &  $ 2.44 + 0.10 $     &  $6.61 \pm 0.24  $       & $ 4.08^{+0.09}_{-0.06} $       \\ 
($R_{500} = 0.96 h^{-1} \,$ Mpc)              & $ 3.79 \pm 0.08  $       &  $  5.53 \pm 0.27  $       &  $ 6.83 \pm 0.30  $      & --- \\
A2597 (500 $\, h^{-1} \,$ kpc)      & $0.902 \pm 0.025 $     &  $ 1.52^{+0.10}_{-0.16 } \,$     &  $5.96^{+0.56}_{-0.34}  $       & $ 45.13^{+2.81}_{-4.75} $       \\ 
($R_{500} = 0.78 h^{-1} \,$ Mpc)              & $ 1.53 \pm 0.07 $       &  $ 2.60^{+0.19}_{-0.29} $       &  $5.91^{+0.57}_{-0.36} $      & --- \\
A3112 (500 $\, h^{-1} \,$ kpc)      & $ 0.974 \pm 0.017  $     &  $ 1.64^{ +0.14}_{- 0.20}  $     &  $5.95^{ + 0.66}_{ - 0.46}   $       & $ 38.38^{+2.30}_{-2.90} $       \\ 
($R_{500} = 0.86 h^{-1} \,$ Mpc)              & $2.03 \pm 0.13  $       &  $ 2.96^{ + 0.29}_{ - 0.39}  $       &  $ 6.84^{+ 0.78}_{ - 0.56}   $      & --- \\
A3158 (500 $\, h^{-1} \,$ kpc)      & $ 1.16 \pm 0.02 $     &  $1.65^{+0.07}_{-0.09}  $     &  $7.06^{+0.26}_{-0.21}  $       & $5.54 \pm 0.24   $       \\ 
($R_{500} = 0.88 h^{-1} \,$ Mpc)              & $  2.50 \pm 0.06 $       &  $ 3.38^{+0.17}_{-0.19} \,$       &  $ 7.38^{+0.35}_{-0.31}  $      & --- \\
A3266 (500 $\, h^{-1} \,$ kpc)      & $1.54 \pm 0.01  $     &  $2.55 \pm 0.10  $     &  $6.06 \pm 0.22  $       & $ 2.49 \pm 0.05  $       \\ 
($R_{500} = 1.06 h^{-1} \,$ Mpc)              & $ 4.69 \pm 0.10  $       &  $ 7.72 \pm 0.37 $       &  $ 6.08 \pm 0.23  $      & --- \\
A3558 (500 $\, h^{-1} \,$ kpc)      & $1.12 \pm 0.02  $     &  $ 1.60 \pm 0.08  $     &  $7.00 \pm 0.32  $       & $ 5.71\pm 0.14  $       \\ 
($R_{500} = 0.88 h^{-1} \,$ Mpc)              & $2.63\pm 0.10   $       &  $ 2.89\pm 0.15  $       &  $9.12\pm 0.42 $      & --- \\
A3571 (500 $\, h^{-1} \,$ kpc)      & $1.42 \pm 0.01  $     &  $2.42 \pm 0.05  $     &  $5.86 \pm 0.10 $       & $ 8.59^{+0.17}_{-0.24}   $       \\ 
($R_{500}=0.98 h^{-1} \,$ Mpc )              & $ 3.33 \pm 0.04  $       &  $ 5.15 \pm 0.16  $       &  $6.48 \pm 0.17  $      & --- \\
A3667 (500 $\, h^{-1} \,$ kpc)      & $ 1.54 \pm 0.05$     &  $2.00 \pm 0.13  $     &  $ 7.73 \pm 0.39$       & $ 4.00 \pm 0.57  $       \\ 
($R_{500} = 0.99 h^{-1} \,$ Mpc)              & $ 4.61 \pm 0.21 $       &  $4.35 \pm 0.45   $       &  $ 10.58 \pm 0.79 $      & --- \\
A3921 (500 $\, h^{-1} \,$ kpc)      & $ 1.11 \pm 0.05 $     &  $ 1.82 \pm 0.30 $     &  $ 6.11 \pm 0.92 $       & $ 6.25^{+0.92}_{ -0.70}    $       \\ 
($R_{500} = 0.94 h^{-1} \,$ Mpc)              & $ 2.90 \pm 0.34  $       &  $ 3.50 \pm 0.65  $       &  $ 8.30 \pm 1.32 $      & --- \\
\hline
\end{tabular}
\end{center}
\caption{Central densities, total masses, and baryon fractions for 11 clusters at
$500 h^{-1} \,$ kpc and $R_{500} \,$.  Errors are $1-\sigma$.}
\label{tbl:mass_distr_2}
\end{table}
\renewcommand{\baselinestretch}{1.6}\small\normalsize \phantom{z}

\clearpage
\section{Discussion and Conclusions}
\label{sec:conclusion}

We have defined an x-ray flux-limited sample of 31 nearby galaxy
clusters and analyzed ROSAT PSPC data on 22 of these.  The primary
focus of our analysis is the quantification of SZE modeling
uncertainties.  Using a suite of Monte Carlo simulations, we find that
on average we predict the inverse Compton optical depth with an
accuracy of $5.4\%$ for the OVRO 5.5-meter telescope.  These
predictions are robust with respect to our cooling-flow modeling strategy and
unaffected by realistic PSPC systematics.  We have also presented
similarly accurate predictions of the inverse Compton optical depth
for the near-future ACBAR experiment.  While somewhat less robust than
the 5.5-meter predictions due to the small angular scales this
telescope samples, these predictions should be more than sufficient
for real-world applications.

We have also confirmed the \citet{Allen_et_al_1993} report of an excess column
density of $N_{H}$ towards A478, but do not find evidence for similar anomalies
in any of 21 other clusters.  There appears to be an excess of soft counts in
the ROSAT PSPC spectra, similar that reported by \citet{Iwasawa_et_al}.

\begin{figure}
\vspace{3.8in}
\includegraphics{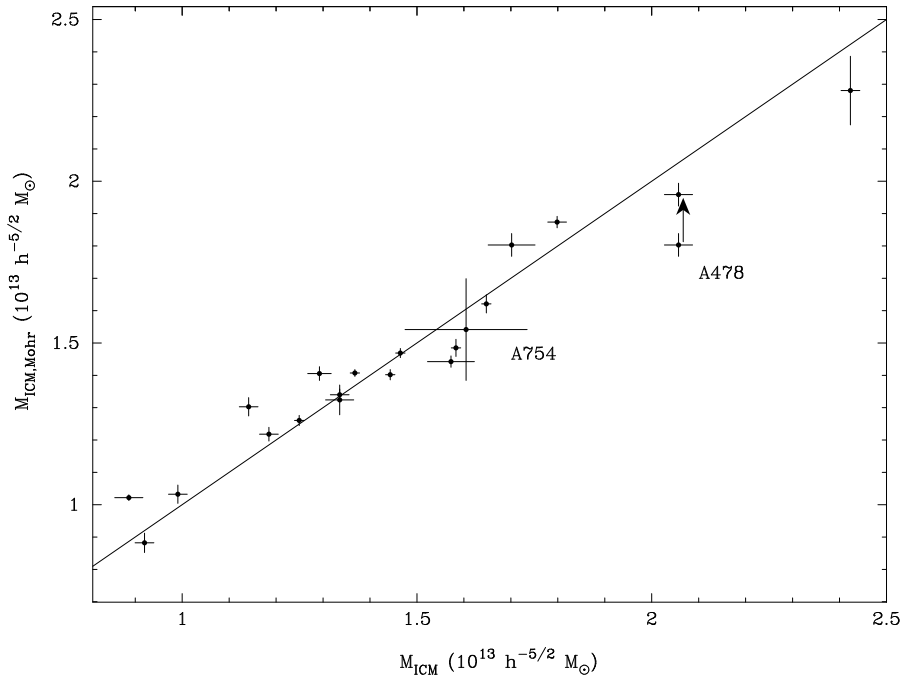}
\caption{A comparison of the ICM masses obtained by \citet{Mohr_et_al_1999} and in this analysis for 20 of our 22 clusters.
Masses are evaluated at $500 \, h^{-1}\,$ kpc.  The arrow shows the effect of correcting the Mohr et al. A478 mass for the observed
excess neutral hydrogen absorption.  A754, the most clearly dynamically disturbed cluster in our sample, is also labelled.}
\label{fig4}
\end{figure}

The mean baryon fraction within $500 \, h^{-1} \, {\rm kpc} \, $ is
found to be $(6.09 \pm 0.17) \, h^{-3/2}\,\% \, ({\rm s.d.} = 0.78
\,h^{-3/2} \,\%) \,$; within $R_{500}$, it is $ (7.02 \pm 0.28 ) \,
h^{-3/2}\,\% \, ( {\rm s.d. } = 1.32\, h^{-3/2}\, \%)$.  For $ h
=0.5$, these are $(17.3 \pm 0.47)\%$ and $(19.8 \pm 0.8)\%$
respectively.  These are consistent with the $10\% - 20\%$ baryon
fractions observed (inside $\,1 \,h^{-1} \, {\rm Mpc}\, $ with
$h=0.5$) by \citet{White_and_Fabian_1995} in analyses of Einstein IPC
data on galaxy clusters.  Our results are also in good agreement with
those of \citet{Mohr_et_al_1999}, who find baryon fractions of $\sim
21\% \,$ within $R_{500}$ for an analysis of the 44 clusters in the
Edge sample with ROSAT PSPC data.  Our mass fractions tend to be
slightly lower on average.  In part this is due to the fact that the
x-ray temperatures used by others in the literature are often biased
low by cooling flow emission; the remainder is due to the difference
in cluster modeling strategies.  For $h \sim 0.7$ our findings also
agree with gas mass fractions determined by \citet{Grego_thesis} on
the basis of interferometric observations of the SZE in distant
clusters.  In a sample of 14 clusters, they find a gas mass fraction
of $\,7.1^{+1.0}_{-1.1} \,h^{-1}\, \%$ within $R_{500}$ (at 68\%
confidence).

While a number of other systematic x-ray cluster analyses exist in the
literature
\citep[e.g.,][]{White_Jones_and_Forman_1997,Mohr_et_al_1999}, ours is
the first to attempt to realistically quantify the effect of the
uncertainties in the x-ray modeling on the predicted SZE decrement for
a given instrument.  Since the analysis of Mohr et al. is closest to
ours in spirit and technique, we have conducted a detailed comparison
of their results to ours.  Figure~\ref{fig4} shows the Mohr et
al. determination of the baryonic mass internal to $500 \, h^{-1} \,
{\rm kpc}$ versus our determination of this quanitity.  The overall
agreement is excellent: the mean mass ratio ($M_{ICM,Mohr\,
et\,al.}/M_{ICM,this\,work}$) is $1.007 \pm 0.016$ (${\rm S.D.} =
7\%$), strongly arguing against any systematic differences in our
analyses.  One of the most significant outliers is A478, for which
Mohr et al.  have used the anomalously low Galactic value for $N_H$;
the effect of correcting for this is shown by an arrow in
Figure~\ref{fig4}.  The scatter between our results, however, is
somewhat larger than the $\sim 4\%$ scatter which is expected.  This
may be in part due to different strategies for profile modeling,
spectral extraction, and dealing with cooling flows.

We have attempted to directly assess the robustness of our results,
but there are several considerations beyond the scope of this analysis
which could affect our conclusions.  The most significant is the
possibility of substructure in the ICM.  \citet{Mathiesen_et_al_1999}
have studied this in simulations of ROSAT PSPC cluster observations
and find a mean overestimate of the cluster density of $\sim 10\%$.
Since these simulations do not include astrophysically important
mechanisms such as cooling and conduction, it will be important to
address this issue with the current and upcoming x-ray missions
Chandra and XMM, as well as more powerful simulations.  Also of some
concern is the possibility of large-scale temperature gradients in the
cluster.  While this will not significantly affect the baryonic mass
models we present here, it will affect thermal SZE predictions and the
inferred total masses.  

For consistency with Myers et al. and many other authors we have
assumed $q_o=0.5$. There is increasing evidence, however, that this
may be wrong \citep[for a summary of the evidence see][]{triangle}.
Due to the $h^{-5/2}$ dependence of the baryonic mass on the distance
scale, these results will be most affected by any errors in the
cosmology we assume; the inverse Compton optical depths and central
densities will be least affected.  We have recomputed the sample
average of the baryonic mass, baryonic mass fraction, and inverse
Compton optical depth for two currently viable cosmologies: an open
($\Omega_m=0.3$) model and a closed $\Lambda$ ($\Omega_m=0.3,
\Omega_{\Lambda} = 0.7$) model.  For the closed $\Lambda$ model, the
sample average baryonic mass is increased by 8.8\%, the baryonic mass
fraction by 5.1\%, and the inverse Compton optical depth by 1.7\%.
For the open model, the baryonic mass is increased by 3.0\%, the
baryonic mass fraction by 1.8\%, and the inverse Compton optical depth
by only 0.6\%.  Clearly this translation must be done on a
cluster-by-cluster basis for a comparison of our $\tau$ predictions
with SZ data.

In paper II we will use the models presented in this work, together
with improved ASCA temperatures, to obtain a measurement of $H_o$ from
SZE measurements conducted with the OVRO 5.5-meter telescope.  The
wider implications for cosmology of this work will also be discussed
therein.

We acknowledge helpful discussions with Maxim Markevitch, Joe Mohr,
Steve Snowden, and Alexey Vikhlinin; we also thank Alexey Vikhlinin
for the use of his software.  We thank Patricia Udomprasert for help
in some of the data reduction.  BSM was supported for part of the
duration of this work by the Zacheus Daniels fund at the University of
Pennsylvania; STM was supported by an Alfred R. Sloan fellowship at the
University of Pennsylvania.

\appendix

\section{Cluster Radial Profiles}
\label{app:profiles}

\begin{figure}
\vspace{8in}
\includegraphics{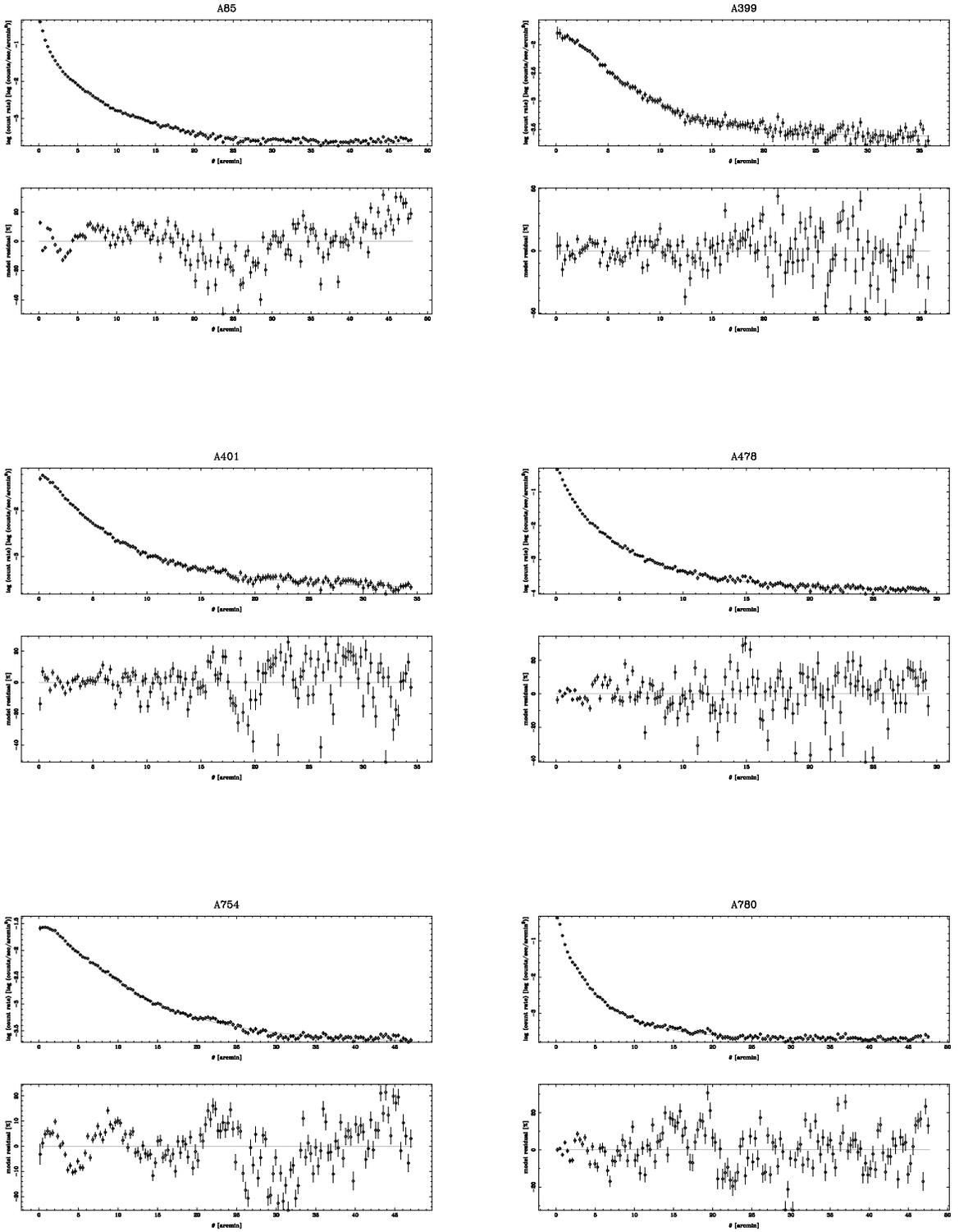}
\caption{Radial profiles for A85, A399, A401, A478, A754, and A780.  The x-axis
is in units of arcminutes and the y-axis the log of the azimuthally averaged count
rate in units of ${\rm counts/sec/arcmin^2}$.}
\label{fig5}
\end{figure}

\begin{figure}
\vspace{8in}

\includegraphics{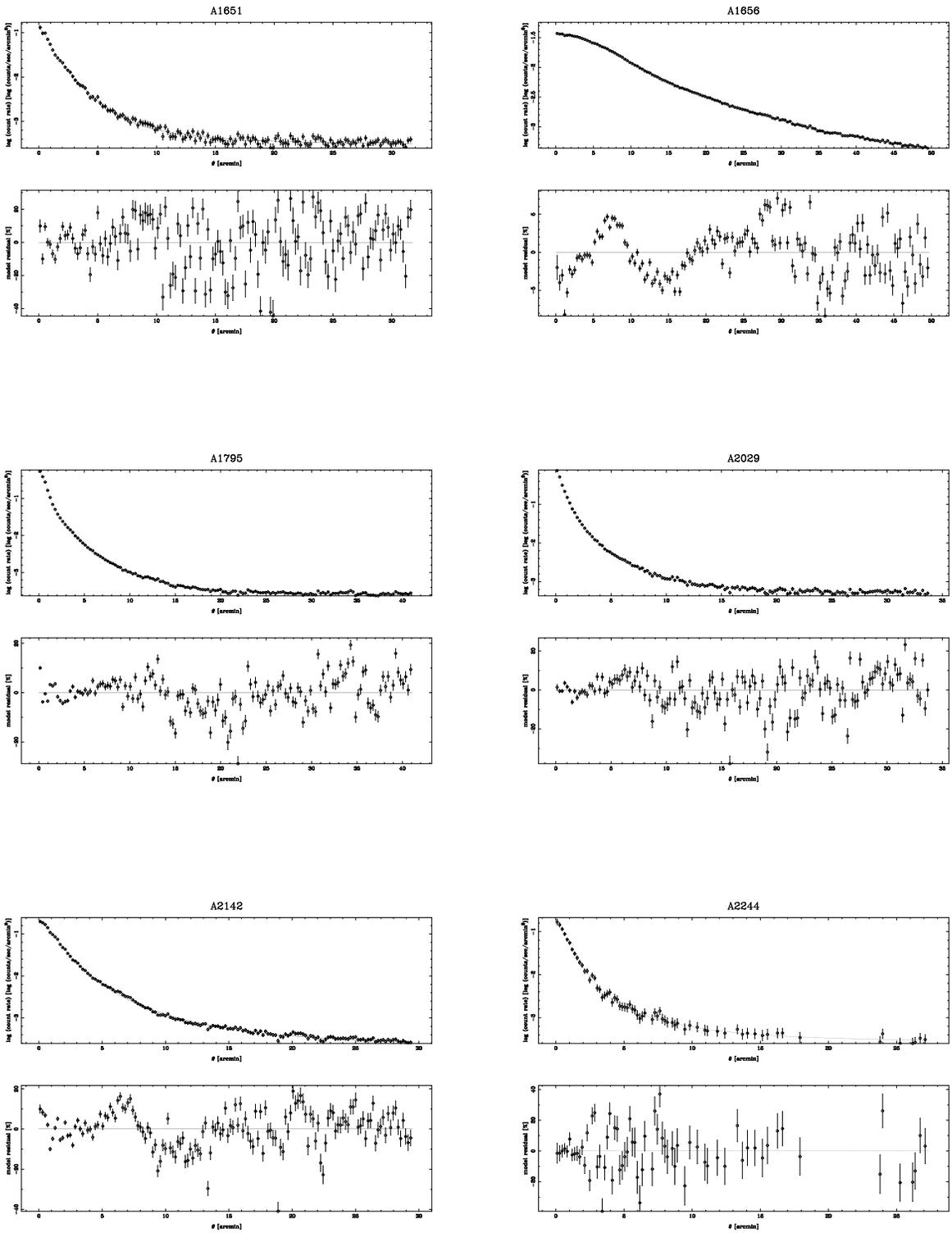}
\caption{Radial profiles for A1651, A1656, A1795, A2029, A2142, and A2244.  The x-axis
is in units of arcminutes and the y-axis the log of the azimuthally averaged count
rate in units of ${\rm counts/sec/arcmin^2}$.}
\label{fig6}
\end{figure}

\begin{figure}
\vspace{8in}
\includegraphics{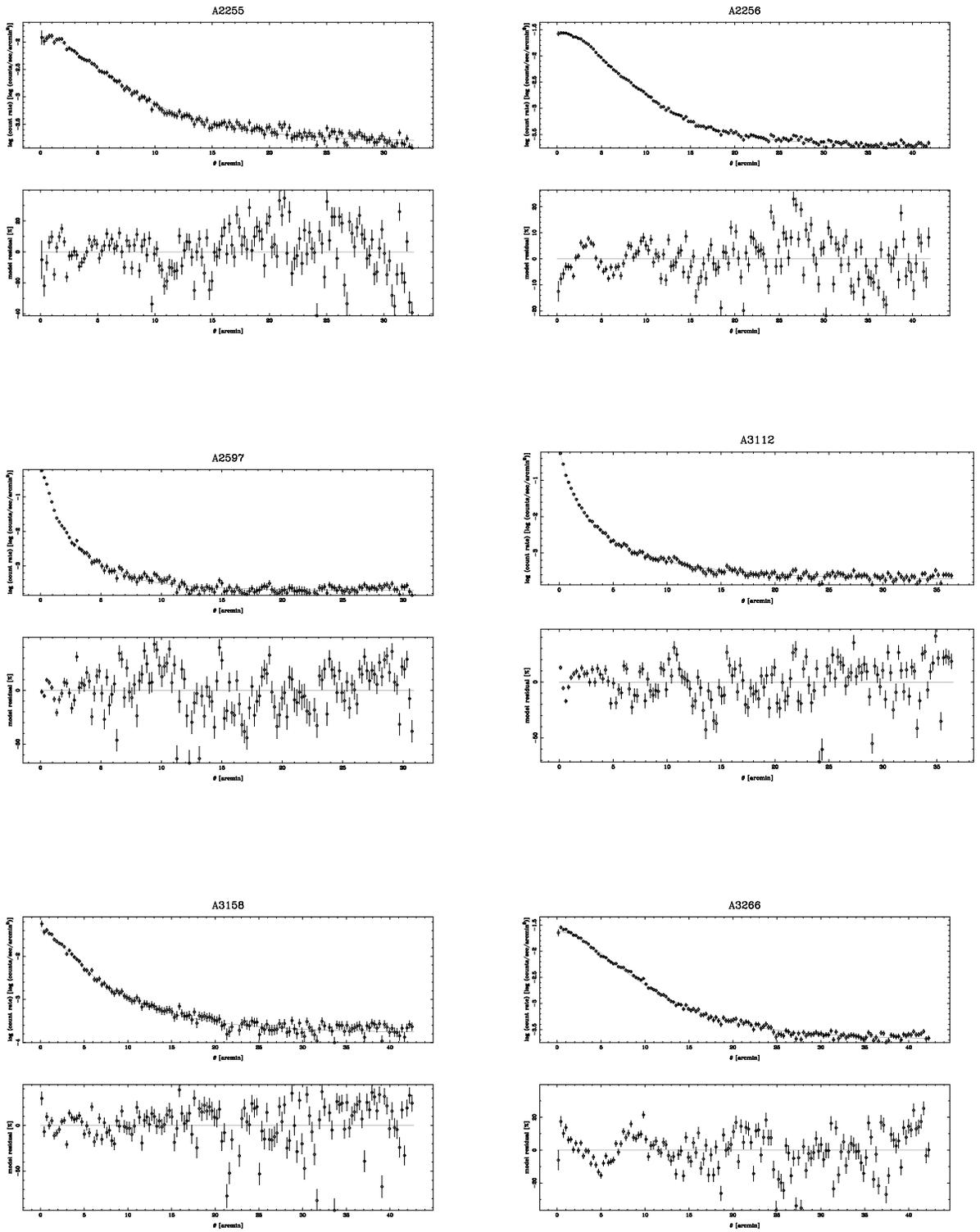}
\caption{Radial profiles for A2255, A2256, A2597, A3112, A3158, A3266.  The x-axis
is in units of arcminutes and the y-axis the log of the azimuthally averaged count
rate in units of ${\rm counts/sec/arcmin^2}$.}
\label{fig7}
\end{figure}

\begin{figure}
\vspace{8in}
\includegraphics{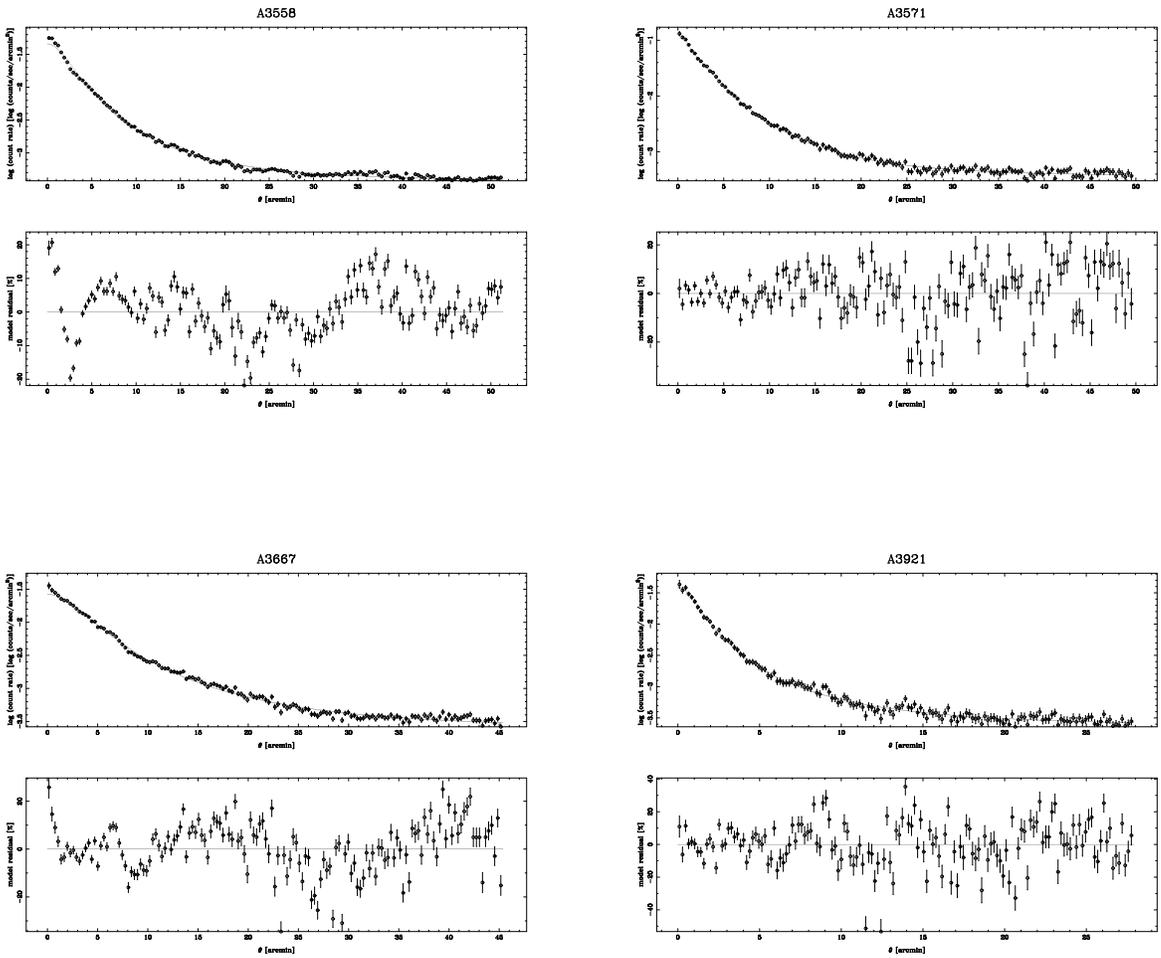}
\caption{Radial profiles for A3558, A3571, A3667, A3921.  The x-axis
is in units of arcminutes and the y-axis the log of the azimuthally averaged count
rate in units of ${\rm counts/sec/arcmin^2}$.}
\label{fig8}
\end{figure}

In this appendix we present the profiles resulting from our analysis
(Figures~\ref{fig5}, \ref{fig6}, \ref{fig7}, and
\ref{fig8}).  The details of this analysis are described in
\S~\ref{sec:dataanalysis}.  Although it is conventional in the
literature to present such fits on a log-log scale, we have chosen to
employ a log-linear scale since this makes the goodness-of-fit at
large radius more readily apparent.

\pagestyle{empty}

\pagestyle{empty}

\pagestyle{empty}

\pagestyle{empty}

\clearpage

%
%
%

%

\end{document}